\documentclass[journal]{IEEEtran} 

\IEEEoverridecommandlockouts

\usepackage{graphicx}
\usepackage{color}

\usepackage{cite}
\usepackage{amsmath}
\usepackage[caption=false]{subfig}

\usepackage{amssymb}
\usepackage{amsfonts}

\usepackage{algorithmic}
\usepackage{algorithm}

\usepackage{bbm}
\usepackage{tikz}
\usepackage{comment}

\newtheorem{proposition}{\bf Proposition}

\newcommand{\CEN}{\textbf{CEN}}
\newcommand{\SYNC}{\textbf{sync-FL}}
\newcommand{\ASYNC}{\textbf{async-FL}}
\newcommand{\FP}{\textbf{FP}}
\newcommand{\QSO}{\textbf{QSO}}
\newcommand{\QSR}{\textbf{QSR}}

\newcommand{\zone}{z}
\newcommand{\ZONE}{Z}
\newcommand{\zoneSet}{\set{\ZONE}}

\newcommand{\vue}{u}
\newcommand{\VUE}{U}
\newcommand{\vueSet}{\set{\VUE}}

\newcommand{\coordinatesTx}{(x_{\vue},y_{\vue})}
\newcommand{\coordinatesRx}{(x_{\vue'},y_{\vue'})}
\newcommand{\pathloss}{\Phi}
\newcommand{\pathlossCoefficient}{\ell}
\newcommand{\pathlossExponent}{c}
\newcommand{\distanceBound}{d_0}
\newcommand{\fading}{\phi}

\newcommand{\rb}{n}
\newcommand{\RB}{N}
\newcommand{\rbSet}{\set{\RB}}

\newcommand{\arrival}{a}
\newcommand{\arrivalAvg}{\overline{\arrival}}

\newcommand{\queue}{q}
\newcommand{\queueTH}{\queue_0}

\newcommand{\queueMAXvar}{m}
\newcommand{\queueAvg}[1]{\expect[\queue_{#1}]}
\newcommand{\outage}{\epsilon}

\newcommand{\rate}{r}
\newcommand{\channel}{h}
\newcommand{\channelVec}{\vect{\channel}}

\newcommand{\fbLength}{D}
\newcommand{\fbErr}{\varepsilon}
\newcommand{\erfcinv}{\text{erfc}^{-1}}

\newcommand{\noiseAlone}{ N_0}
\newcommand{\noise}{\bandwidth \noiseAlone}
\newcommand{\bandwidth}{W}

\newcommand{\txpower}{p}
\newcommand{\txpowerVec}{\vect{\txpower}}
\newcommand{\txpowerMax}{\txpower_0}
\newcommand{\interference}{I}
\newcommand{\interferenceEST}{\tilde{\interference}}

\newcommand{\thresholdMone}{M_1}
\newcommand{\thresholdMtwo}{M_2}

\newcommand{\gevCombined}{\vect{d}}

\newcommand{\loglikelihood}{f^{\gevCombined}}
\newcommand{\loglikelihoodSP}[1]{f^{#1}}

\newcommand{\gpdScale}{\sigma}
\newcommand{\gpdShape}{\xi}
\newcommand{\queuePoT}{M}
\newcommand{\queuePoTset}{\set{M}}
\newcommand{\gpd}{G_{\queuePoT}}
\newcommand{\gpdExponent}[1]{g^{\gevCombined}(#1)}
\newcommand{\gpdCombined}{\vect{d}}
\newcommand{\gpdCombinedFeasible}{\set{D}}
\newcommand{\gpdML}{G^{\gevCombined}_{X}}

\newcommand{\gradient}{\Upsilon}
\newcommand{\localmodel}{( \gradient_{\vue}, \gpdCombined_{\vue}, \BLOCK_{\vue}, \sampleMax_{\vue})}
\newcommand{\globalmodel}{( \gradient, \gpdCombined, \sum_{\vue} \BLOCK_{\vue}, \sampleMax)}

\newcommand{\timeblock}{T}

\newcommand{\si}{\alpha}
\newcommand{\cinr}{\eta}
\newcommand{\sinr}{\gamma}
\newcommand{\dualPower}{\lambda}

\newcommand{\vqQ}{\Psi}
\newcommand{\vqEXq}{A}
\newcommand{\vqEXqtwo}{B}

\newcommand{\lyapunov}[1]{L(#1)}
\newcommand{\lyapunovDrift}{\Delta L_t}
\newcommand{\queueCombined}{\Xi}
\newcommand{\lyapunovBound}{\Delta_0}
\newcommand{\lyapunovConst}{\Delta_{\vue}}
\newcommand{\lyapunovTradeoff}{V}

\newcommand{\block}{k}
\newcommand{\BLOCK}{K}
\newcommand{\blockTH}{\tilde{\BLOCK}}
\newcommand{\timeFederate}{\tilde{t}}

\newcommand{\blocklength}{w}
\newcommand{\sample}{Q}
\newcommand{\sampleSet}{\set{\sample}}
\newcommand{\sampleMax}{\hat{\sample}}
\newcommand{\blockTimeSet}{\set{T}}
\newcommand{\sampleSizeRatio}{\kappa_{\vue}}

\newcommand{\federatedTime}{T_f}
\newcommand{\stepsize}{\delta}

\newcommand{\sizeGradient}{J_{\gradient}}
\newcommand{\sizeParams}{J_{\gpdCombined}}
\newcommand{\sizeQueues}{J_{\sample}}

\newcommand{\txr}{\text{vTx-vRx}}%

\newcommand*{\myfigfactor}{0.95}
\newcommand*{\myfigfactorX}{.9}
\newcommand*{\myfigfactorx}{.9}
\newcommand*{\myfigfactorxx}{0.238}

\DeclareMathOperator*{\argmax}{arg\,max}
\DeclareMathOperator*{\argmin}{arg\,min}

\newcommand{\optMinimize}{\min}

\newcommand{\subjectTo}{\text{s.t.}}

\newcommand{\expect}{\mathbb{E}\,}
\newcommand{\probability}{\text{Pr}}

\newcommand{\naturalset}{\mathbb{N}}

\newcommand{\vect}{\boldsymbol}

\newcommand{\seta}[1]{1,\dots,#1}
\newcommand{\set}[1]{\mathcal{#1}}
\newcommand{\setSize}[1]{|#1|}

\newcommand{\vectx}{\vect{x}}

\newcommand{\one}{\mathbf{1}}
\newcommand{\zero}{\mathbf{0}}

\newcommand{\indictsimp}[1]{\mathbbm{1}_t}

\newcommand{\transpose}{^\dag}
\newcommand{\optimal}{^\star}

\newcommand{\grad}[2]{\nabla_{#1}#2}
\newcommand{\daba}[2]{\frac{\partial #1}{\partial #2}}

\newcommand{\norm}[2]{\| #1 \|_{#2}}

\newcommand{\myeq}{\mathrel{\overset{\makebox[0pt]{\mbox{\normalfont\tiny\sffamily $a.s.$}}}{=}}}
\newcommand{\myleq}{\mathrel{\overset{\makebox[0pt]{\mbox{\normalfont\tiny\sffamily $a.s.$}}}{\leq}}}

\usepackage{forloop}
\newcounter{loopcntr}
\newcommand{\rpt}[2][1]{%
	\forloop{loopcntr}{0}{\value{loopcntr}<#1}{#2}%
}

\newcommand{\subgroup}[1]%
{\rlap{\smash{%
	\newcount\cnt%
	\cnt \numexpr#1\relax%
	\advance\cnt -1\relax%
	$\tabcolsep=.1em\begin{tabular}[t]{|l}\multicolumn{1}{l}{}\\%
	\rpt[\cnt]{\\}
	\\\hline\end{tabular}$%
}}}

\newcommand{\zonedef}{group}

\newcounter{myRefCount}

\newcounter{reviewerQ}

{\noindent\ignorespaces\refstepcounter{reviewerQ}\itshape \emph{\thereviewerQ}.%
	 }%
{\vspace{5mm}}

\allowdisplaybreaks

\begin{document}

\title{%
	Distributed Federated Learning for Ultra-Reliable Low-Latency Vehicular Communications
}

\author{
	\IEEEauthorblockN{
		Sumudu Samarakoon, \IEEEmembership{Member, IEEE},
		Mehdi Bennis, \IEEEmembership{Senior~Member, IEEE},
		Walid Saad, \IEEEmembership{Fellow, IEEE},
		and M\'{e}rouane Debbah, \IEEEmembership{Fellow, IEEE}
		\\}
	\thanks{
		%
		Preliminary results of this work is published in \cite{z_previous_work}.
	}
	\thanks{
		Sumudu Samarakoon is with Centre for Wireless Communication, University of Oulu, FI-90014 Oulu, Finland (email: sumudu.samarakoon@oulu.fi)
	}
	\thanks{
		Mehdi Bennis is with Centre for Wireless Communication, University of Oulu, FI-90014 Oulu, Finland and also with the Department of Computer Science and Engineering, Kyung Hee University, Seoul 130-701, South Korea (email: mehdi.bennis.fi)
	}
	\thanks{
		Walid Saad is with Wireless@VT, Bradley Department of Electrical and Computer Engineering, Virginia Tech, Blacksburg, VA 24061 USA (email: walids@vt.edu)
	}
	\thanks{
		M\'{e}rouane Debbah is with Mathematical and Algorithmic Sciences Lab, Huawei France Research and Development, 92100 Boulogne-Billancourt, France (email: merouane.debbah@huawei.com)
	}
}

\maketitle
\nopagebreak[4]
\begin{abstract}

In this paper, the problem of joint power and resource allocation (JPRA) for ultra-reliable low-latency communication (URLLC) in vehicular networks is studied. 
Therein, the network-wide power consumption of vehicular users (VUEs) is minimized subject to high reliability in terms of probabilistic queuing delays. 
Using extreme value theory, a new reliability measure is defined to characterize extreme events pertaining to vehicles' queue lengths exceeding a predefined threshold. 
To learn these extreme events,
assuming they are independently and identically distributed over VUEs,
a novel distributed approach based on federated learning (FL) is proposed to estimate the tail distribution of the queue lengths. 
Considering the communication delays incurred by FL over wireless links, Lyapunov optimization is used to derive the JPRA policies enabling URLLC for each VUE in a distributed manner. 
The proposed solution is then validated via extensive simulations using a Manhattan mobility model. 
Simulation results show that FL enables the proposed method to estimate the tail distribution of queues with an accuracy that is close to a centralized solution with up to 79\% reductions in the amount of exchanged data. 
Furthermore, the proposed method yields up to 60\% reductions of VUEs with large queue lengths, while reducing the average power consumption by two folds, compared to an average queue-based baseline.

\end{abstract}

\section{Introduction}\label{sec:introduction}

Providing efficient vehicle-to-vehicle (V2V) communications is a necessary stepping stone for enabling autonomous and intelligent transportation systems (ITS) \cite{tech:5gcar,jnl:shah18,jnl:liu18,jnl:zeng18,pap:ikram17}.
V2V communications can extend drivers' field of view, thus enhancing traffic safety and driving experience, while enabling new transportation features such as platooning, real-time navigation, collision avoidance, and autonomous driving \cite{jnl:zeng18,tech:5gcar}.
However, the performance of emerging transportation applications heavily rely on the availability of V2V communication links with extremely low errors and delays. 
In this regard, achieving ultra-reliable low-latency communication (URLLC) for V2V networks is necessary for realizing the vision of intelligent transportation~\cite{tech:5gcar,jnl:saad19}.
The modeling of URLLC has focused on different system design aspects such as guaranteeing the signal-to-interference ratio, data rate, over-the-air/queuing latency, connectivity, age-of-information (AoI), and decoding probability \cite{jnl:Popovski17,pap:Pan19}.
Since over-the-air latency and queuing latency are coupled, ensuring low queuing latency is required to achieve the much coveted target end-to-end latency of 1\,ms. 
This, in turn, necessitates efficient radio resource management (RRM) techniques \cite{tech:3gpp,jnl:mozaffari16,pap:ikram17,jnl:mozaffari19}.
Furthermore, the increased energy consumption and its negative impact on the environment due to the large number of vehicles in modern transportation system, and improving energy-efficiency/energy savings need to be addressed within RRM in V2V communications \cite{jnl:zhou18,jnl:kumar14}.

Several existing RRM techniques have been proposed for enabling ultra-reliable low-latency vehicular communications while factoring in several challenges such as rate maximization, delay minimization, improving energy-efficiency, energy saving, and vehicle clustering/platooning \cite{jnl:zeng18,pap:ikram17,jnl:liu18,	jnl:zhou18,jnl:kumar14,pap:dong16,pap:sun14,pap:perfecto17,pap:ikram16,pap:zhou18,jnl:TLiu17,jnl:mei18,z_previous_work}.
In \cite{jnl:zeng18}, the performance of vehicular platooning is optimized while jointly considering the delay of the wireless network and the stability of the vehicle's control system.
By grouping vehicles into clusters, the work in \cite{pap:ikram17} minimizes the total transmission power in a vehicular network while considering queuing latency and reliability.
In \cite{jnl:zhou18}, an energy-efficient resource allocation algorithm is proposed for cooperative V2V communication systems.
The work in \cite{jnl:kumar14} proposed an energy saving sleep mode strategy for access points serving motorway vehicular traffic.
The problem of vehicle network clustering is studied in \cite{pap:dong16} to reduce the power consumption of V2V communications.
In \cite{pap:sun14}, a joint resource allocation and power control algorithm is proposed to maximize the V2V sum rate.
The authors in \cite{pap:perfecto17} optimize the beam alignment and scheduling among vehicles to reduce the V2V communication delays.
In \cite{pap:ikram16}, the tradeoff between service delay and transmission success in V2V communications is optimized.
The URLLC aspects of this prior art \cite{jnl:zeng18,pap:ikram17,jnl:zhou18,jnl:kumar14,pap:dong16,pap:sun14,pap:perfecto17,pap:ikram16,pap:zhou18} are captured by either improving the average latencies or imposing a probabilistic constraint to maintain small queue lengths.
Although such a probabilistic constraint on the queue length improves network reliability, it fails to control rare events in which large queue lengths occur with low probability, i.e., the tail distribution of queue lengths.
As a result, if the network relies on these existing schemes, some of the vehicular users (VUEs) may experience unacceptable latencies yielding degraded performance \cite{jnl:bennis18,jnl:liu18,pap:liu07,jnl:MOURADIAN16,pap:wei02}.

In practice, to enable a truly URLLC experience, it is imperative to model and capture extreme, low probability events.
To this end, \emph{extreme value theory} (EVT), a powerful tool from statistics that characterizes the occurrences of extreme, low probability events instrumental in enabling URLLC~\cite{book:EVTfinkenstad03}. 
In \cite{pap:liu07}, EVT is used to model the distributions of data rates exceeding a threshold for few traffic traces and then, the accuracy of the analytical model is evaluated using simulations.
The work in \cite{jnl:MOURADIAN16} studies the statistical distributions of inter-beacon delays in safety applications for vehicular adhoc networks (VANETs) using EVT.
The authors in \cite{pap:wei02} use EVT to model the peak distribution of the orthogonal frequency division multiplexing envelope while characterizing the variations in peak-to-average-power ratios.
The work in \cite{jnl:liu18} employs EVT to characterize the statistics of maximal queue length so as to control the worst-case latency of V2V communication links therein.
Characterizing the distribution of extreme events using EVT, i.e., determining the location, shape, and scale parameters of the tail distribution, in the above works necessitates the acquisition of sufficient samples capturing extreme events.
Depending on the network size and the quality of the communication within the network, the process of gathering samples over the network may introduce unacceptable overheads that are not investigated in the aforementioned works.
In a real-time system such as a V2V communication network, VUEs may have access to limited number of queue length samples (particularly those that are locally in excess over a high threshold) and hence they are unable to estimate the tail distribution of the network-wide queue lengths.
Therefore, roadside units (RSUs) can assist in gathering samples over the network at a cost of additional data exchange overheads.
Furthermore, due to the resource limitations available for V2V communication, VUEs may be unwilling to allocate their resources to share their individual queue state information (QSI) with an RSU and other VUEs.
This shortcoming warrants a collaborative learning model that does not rely on sharing individual QSI.

Recently, \emph{federated learning} (FL) was proposed as a decentralized learning technique where training data is distributed (possibly unevenly) across learners, instead of being centralized \cite{jnl:jakub16,jnl:smith17}.  
FL allows each learner to derive a set of local learning parameters from the available training data, referred to as \emph{local model}.
Instead of sharing the training data, learners share their local models with a central entity, which in turn does model averaging then sharing a \emph{global model} with the learners. 
In \cite{jnl:jakub16}, the applicability of several existing algorithms for FL are studied and a novel algorithm is proposed to handle the sparse data available at individual learners.
The means of minimizing the communication cost by sharing a reduced number of parameters of FL models are discussed in \cite{pap:jakub16}.
In \cite{jnl:smith17}, FL is used to develop distributed learning models for multiple related tasks simultaneously, referred to as multi-task learning.
The recent work in \cite{pap:nishio18}, proposes a new FL protocol that solves a client selection problem with resource constraints in mobile edge computing. 
Our prior work in \cite{z_previous_work} proposes a distributed FL-based algorithm for VUEs based on a maximum likelihood estimation (MLE).
However, this prior work does not consider sharing wireless resources for FL and V2V communications, whereby the impact of FL over shared wireless resources on V2V URLLC is not investigated.
To the best of our knowledge, with the exception of \cite{z_previous_work}, no work has studied the use of federated learning in the context of URLLC.

The main contribution of this paper is to propose a distributed, FL-based, joint transmit power and resource allocation framework for enabling ultra-reliable and low-latency vehicular communication.
We formulate a network-wide power minimization problem while ensuring low latency and high reliability in terms of probabilistic queue lengths. 
To model reliability, we first obtain the statistics of the queue lengths exceeding a high threshold by using the EVT notion of a \emph{generalized Pareto distribution} (GPD)~\cite{book:EVTfinkenstad03}.
Using the statistics of the GPD, we impose a local constraint on extreme events pertaining to queue lengths exceeding a predefined threshold for each VUE.
Here, the characteristic parameters of the GPD are known as scale and shape, which are obtained by using the MLE.
In contrast to the classical MLE design which requires a central controller (e.g., RSU) to collect samples of queue lengths exceeding a threshold from all VUEs in the network, using FL every vehicle builds and shares its own local model (two gradient values) with the RSU.
The RSU aggregates the local models, does model averaging across vehicles, and feeds back the global model to VUEs.
Leveraging different time scales, using our proposed approach, each VUE learns its GPD parameters locally in a short time scale while the model averaging (global learning) takes place in a longer time scale. 
	Here, an assumption of independently and identically distributed queue length samples exceeding the threshold over all VUEs is imposed to ensure accurate GPD parameter estimation using FL.
In our model, we take into account the communication overheads of URLLC due to the model exchange over shared wireless resources.
Then, we propose a distributed algorithm that allows all VUEs to simultaneously learn the GPD parameters using FL.
To further reduce the overhead due to the need of synchronization and simultaneous model sharing, next we develop an asynchronous FL algorithm for MLE that allows VUEs to model and independently learn the tail distribution of queue lengths in a distributed manner. 
Finally, Lyapunov optimization is used to decouple and solve the network-wide optimization problem per VUE.
Simulation results show that the proposed solutions estimate the GPD parameters very accurately compared to a centralized learning module and yields significant gains in terms of reducing the number of VUEs with large queue lengths while minimizing power consumption.
For dense systems with 100 VUE pairs, the proposed solution yields about $60.9\%$ reduction of VUEs with large queue lengths by reducing the power consumption by two folds, compared to a baseline model that controls the reliability using a probabilistic constraint on average queue lengths.
Furthermore, $28.6\%$ and $33.2\%$  reductions in averages and fluctuations of extreme queue lengths, respectively, can be seen in the proposed solution compared to the aforementioned baseline.

The rest of the paper is organized as follows.
Section \ref{sec:system_model} describes the system model and the network-wide power minimization problem.
The distributed solution based on EVT and Lyapunov optimization is presented in Section \ref{sec:distributed_solution}.
In Section \ref{sec:federated_learning}, estimation of the extreme value distribution using FL and the cost of enabling FL for both synchronous and asynchronous approaches are discussed.
Section \ref{sec:results} evaluates the proposed solution by extensive set of simulations.
Finally, conclusions are drawn in Section \ref{sec:conclusion}.
\section{System Model and Problem Definition}\label{sec:system_model}

\begin{figure}[!t]
	\centering
	\includegraphics[width=\myfigfactor\linewidth]{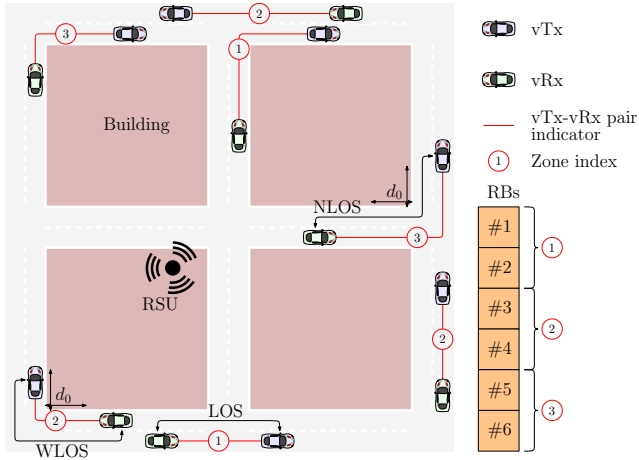}
	\caption{Simplified illustration of the system model containing \txr~pairs within their \zonedef indexes and RB allocation over {\zonedef}s.}
	\label{fig:system_model}
\end{figure}

Consider a vehicular network consisting of a set $\vueSet$ of $\VUE$ communicating VUE pairs, using an RSU that allocates a set $\rbSet$ of resource blocks (RBs) over a partition of the network $\zoneSet$ defined as {\zonedef}s.
Here, a \emph{\zonedef} consists of VUE pairs that can reuse the same RBs with low-to-no interference on one another. 
The RSU allocates RBs orthogonally across the {\zonedef}s to reduce the interference among nearby VUE pairs. 
Hence, a VUE pair $\vue$ is only allowed to use the subset $\rbSet_{\zone(t,\vue)} \subseteq \rbSet$ of RBs allocated to its corresponding \zonedef $\zone(t,\vue)$ at time $t$.
We denote the VUE transmitter (vTx) and receiver (vRx) that belong to VUE pair $\vue$ by vTx $\vue$ and vRx $\vue$, hereinafter.
An illustration of our system model is presented in Fig. \ref{fig:system_model}.

Let $\txpowerVec_{\vue}(t) = [\txpower_{\vue}^{\rb}(t)]_{\rb\in\rbSet_{\zone(t,\vue)}}$ and $\channelVec_{\vue\vue'}(t)=[\channel_{\vue\vue'}^{\rb}(t)]_{\rb\in\rbSet_{\zone(t,\vue)}}$ be, respectively, the transmit power vector of vTx $\vue$, and the channel gain vector between vTx $\vue$ and vRx $\vue'$ over the subset of allocated RBs at time $t$.
Depending on whether the vTx and vRx are located in the same lane or separately in perpendicular lanes, the channel model is categorized into three types:
\emph{i) Line-of-sight} (LOS): both vTx $\vue$ and vRx $\vue'$ are located in the same lane,
\emph{ii) Weak-line-of-sight}  (WLOS): vTx $\vue$ and vRx $\vue'$ are in perpendicular lanes and at least one of them is located at a distance of	 no more than $\distanceBound$ from the corresponding intersection,
and
\emph{iii) Non-line-of-sight}  (NLOS), otherwise.
Let $\coordinatesTx$ and $\coordinatesRx$ be the Cartesian coordinates of vTx $\vue$ and vRx $\vue'$, respectively.
The channel $\channel_{\vue\vue'} = \fading_{\vue\vue'} \pathloss_{\vue\vue'}$ includes a fast fading component $\fading_{\vue\vue'}$ following a Rayleigh distribution with a unit scale parameter 
for LOS, a Nakagami-m distribution with $\text{m}=1.41$ and a unit scale for both WLOS and NLOS,
 in addition to
a path loss model $\pathloss_{\vue\vue'}$ for urban areas at 5.9\,GHz carrier frequency~\cite{pap:mangel11}:
\begin{equation}\label{eqn:path_loss_model}
	\pathloss_{\vue\vue'} = 
	\begin{cases}
		\pathlossCoefficient \norm{\coordinatesTx-\coordinatesRx}{2}^{-\pathlossExponent} & \text{for LOS}, \\
		\pathlossCoefficient \norm{\coordinatesTx-\coordinatesRx}{1}^{-\pathlossExponent} & \text{for WLOS}, \\
		\pathlossCoefficient' ( |x_{\vue}-x_{\vue'}| \cdot |y_{\vue}-y_{\vue'}| )^{-\pathlossExponent} & \text{for NLOS},
	\end{cases}
\end{equation}
where $\norm{\vectx}{l}$ is the $l$-th norm of vector $\vectx$, $\pathlossExponent$ is the path loss exponent, and the path loss coefficients $\pathlossCoefficient$ and $\pathlossCoefficient'$ satisfy $\pathlossCoefficient' < \pathlossCoefficient (\frac{\distanceBound}{2})^{\pathlossExponent}$.
The transmission rate between the \txr~pair $\vue$ is given by,
\begin{equation}\label{eqn:rate_vue_pair}
	\rate_{\vue}(t) 
	= \textstyle \sum\limits_{\rb\in\rbSet_{\zone(t,\vue)}} \rate_{\vue}^{\rb}(t) 
	= \textstyle \sum\limits_{\rb} \bandwidth 
	\log_2 \left( 1 + \frac{ \channel_{\vue\vue}^{\rb}(t) \txpower_{\vue}^{\rb}(t) }{ \interference_{\vue}^{\rb}(t) + \noise }\right),
\end{equation}
where $\interference_{\vue}^{\rb}(t) = \sum_{\vue'\in \vueSet\setminus\{\vue\} }  \channel_{\vue'\vue}^{\rb}(t) \txpower_{\vue'}^{\rb}(t)$ is the interference from other vTxs, $\bandwidth$ is the bandwidth of each RB, and $\noiseAlone$ is the noise power spectral density.
At each time $t$, $\arrival_{\vue}(t)$ data bits are randomly generated with a mean of $\arrivalAvg_{\vue}$ at vTx $\vue$ that must be delivered to its corresponding vRx.
Thus, at the vTx, a data queue is maintained and has the following dynamics:
\begin{equation}\label{eqn:queue_dynamics}
	\queue_{\vue}(t+1) = [\queue_{\vue}(t) + \arrival_{\vue}(t) - \rate_{\vue}(t)]^+,
\end{equation}
where $[x]^+ = \max(x,0)$.

The number of vehicles is expected to grow continuously, in which improving energy efficiency and saving energy in vehicular networks is a key requirement.
Our goal is therefore to minimize the network-wide power consumption while ensuring URLLC.
Considering use cases such as information exchange on blind-spots or sequences of future actions (turning, parking, slowing/speeding) based on the traffic ahead, it is important to optimize decision making taking into account the queue length and its tail distributions \cite{bai2018dynamic,pap:ikram17,jnl:liu18}.
In this view, 
here, reliability is achieved by guaranteeing queue stability for each vTx while keeping outages below a predefined threshold, i.e., the probability that the queue length exceeding a threshold $\queueTH$ is below a certain probability $\outage$. 
The reliability conditions can now be formally defined as:
\begin{gather}
	\label{eqn:queue_stability}
	\queueAvg{\vue} \myeq \textstyle \lim_{\timeblock \to \infty} \frac{1}{\timeblock}\sum_{t=1}^{\timeblock} \queue_{\vue}(t) < \infty \qquad \forall \vue\in\vueSet, \\
	\label{eqn:queue_reliability}
	\probability( \queue_{\vue}(t) \geq \queueTH) \leq \outage \qquad \forall \vue\in\vueSet, \forall t.
\end{gather}

Note that the above reliability constraints cannot cope with 
the extreme cases in which the queue lengths  $\queue_{\vue}(t) > \queueTH$ 
that occur with a probability below $\outage$.
Such extreme cases essentially correspond to the worst case network queuing latency (as well as end-to-end latency \cite{tech:3gpp,jnl:mozaffari16,pap:ikram17}) which are a key determinant of the URLLC performance and, hence, must be properly addressed.
In this regard,
the set of queue length samples exceeding the threshold $\queueTH$ over the network is defined as $\queuePoTset = \{ (\queue-\queueTH) | \queue>\queueTH, \queue\in\{\queue_{\vue}(t)\}_{\vue\in\vueSet}^t \}$.
Then, a sample of an extreme event, i.e., queue length exceeding $\queueTH$, is $\queuePoT\in\queuePoTset$.
By imposing the following constraints for all VUEs,
\begin{gather}
\label{eqn:constraintNW_extremes_bounds}
	\textstyle \lim\limits_{\timeblock \to \infty} \sum_{t=1}^{\timeblock} \big( \queue_{\vue}(t) - \queueTH \big)\indictsimp{\queue_{\vue}(t)} / \sum_{t=1}^{\timeblock} \indictsimp{\queue_{\vue}(t)} \myleq \expect[\queuePoT], \\
\label{eqn:constraintNW_extremes_bounds_2}
	\textstyle \lim\limits_{\timeblock \to \infty} \sum_{t=1}^{\timeblock} \big( \queue_{\vue}(t) - \queueTH \big)^2 \indictsimp{\queue_{\vue}(t)} / \sum_{t=1}^{\timeblock} \indictsimp{\queue_{\vue}(t)} \myleq \expect[\queuePoT^2],
\end{gather}
each VUE $\vue$ can better control the fluctuations of its queue and maintain its extreme values below the desired threshold.
Here, $\indictsimp{x}$ is an indicator function with $\indictsimp{x}=1$ when $\queue(t)>\queueTH$, and $\indictsimp{x}=0$, otherwise.
Note that the benefits of imposing $\expect[\queuePoT]$ and $\expect[\queuePoT^2]$ as targets of \eqref{eqn:constraintNW_extremes_bounds} and \eqref{eqn:constraintNW_extremes_bounds_2} include:
\emph{i)} The targets are naturally generated within the system which can prevent infeasibility and instability under predefined targets,
\emph{ii)} The chosen targets allow vehicular users with larger queue lengths to exploit more transmit power compared to ones with short queue lengths allowing to control interference,
and
\emph{iii)} In consequence, the tail distributions of all vehicles converge to identical distributions, which is essential for our analysis.
Next, we can now formally pose our network-wide power minimization problem: 
\begin{subequations}\label{eqn:optimization_NW}
\begin{eqnarray}
	\label{eqn:objectiveNW_min_power}
	\underset{[\txpowerVec_{\vue}(t)]_{\forall\vue\in\vueSet}^{\forall t}}{\optMinimize} && \textstyle \lim_{\timeblock \to \infty}  \frac{1}{\timeblock}\sum_{t=1}^{\timeblock}   \sum_{\vue\in\vueSet} \one\transpose \txpowerVec_{\vue}(t)  \\
	\label{eqn:constraintNW_reliability_modified}
	\subjectTo && \eqref{eqn:queue_dynamics}\text{-}\eqref{eqn:constraintNW_extremes_bounds_2}, \\
	\label{eqn:constraintNW_control_domain}
	&& \txpowerVec_{\vue}(t) \succcurlyeq \zero, \, \one\transpose\txpowerVec_{\vue}(t) \leq \txpowerMax \quad \forall \vue\in\vueSet.
\end{eqnarray}
\end{subequations}
Here, \eqref{eqn:constraintNW_reliability_modified} ensures queue dynamics and reliability while controlling the worst-case latency over all VUEs and $\txpowerMax$ is the transmit power budget of a VUE.
Solving \eqref{eqn:optimization_NW} to obtain the optimal transmission control policy over time is challenging due to two reasons:
\emph{i)} A decision at time $t$ relies on future network states,  
and
\emph{ii)} The characteristics of the distribution of $\queuePoT$ for constraint \eqref{eqn:constraintNW_extremes_bounds} are unavailable.
Moreover, solving \eqref{eqn:optimization_NW} using a centralized approach requires exchanging channel state information (CSI) and QSI over the whole network resulting in unacceptable signaling overheads.
Therefore, a distributed solution that requires minimal coordination within the vehicular network is needed.

\section{Proposed Distributed Framework using EVT and Lyapunov Optimization}\label{sec:distributed_solution}

Developing a distributed solution for solving \eqref{eqn:optimization_NW} requires decoupling the optimization problem over VUE pairs.
Therefore, next, we propose new solutions to decouple the objective function \eqref{eqn:objectiveNW_min_power} and the constraints \eqref{eqn:constraintNW_extremes_bounds} and \eqref{eqn:constraintNW_extremes_bounds_2} based on the statistics of queue lengths exceeding $\queueTH$ over the vehicular network.

\subsection{Modeling Extreme Queue Lengths Using Extreme Value Theory}\label{sec:evt}

The samples of queue lengths exceeding the threshold
	$\queuePoT\in\queuePoTset$
are seen as extreme statistics of the system,
and can be characterized using EVT.
Assume that the individual queues at a given time $[\queue_{\vue}(t)]_{\vue\in\vueSet}$ are samples of independent and identical distributions (IID) and the queue threshold $\queueTH$ is large.
Then, the distribution of $\queuePoT$ can be modeled as a GPD using \cite[Theorem 3.2.5]{book:EVTfinkenstad03}.
This fundamental EVT result mainly shows that, as 
$\queueTH \to \sup \{ q | \probability(\queuePoT > q ) > 0 \}$, 
the conditional probability distribution of 	$\queuePoT\in\queuePoTset$
is given by,
\begin{equation}\label{eqn:GPD}
\gpd^{\gpdCombined} (\queueMAXvar) = 
\begin{cases}
\frac{1}{\gpdScale}(1 + \gpdShape\queueMAXvar/\gpdScale)^{-1-1/\gpdShape} & \text{for~} \gpdShape \neq 0, \\
\frac{1}{\gpdScale} e^{-\queueMAXvar/\gpdScale}  & \text{for~} \gpdShape = 0,
\end{cases}
\end{equation}
with $\gpdCombined = [\gpdScale,\gpdShape]$, and $\gpdShape$ and $\gpdScale(>0)$ are called the shape and scale parameters, respectively.
Here, $\queueMAXvar \geq 0$ if $\gpdShape \geq 0$ while $0 \leq \queueMAXvar \leq -\gpdScale/\gpdShape$ when $\gpdShape < 0$.
Moreover, $\expect[\queuePoT]$ and $\expect[\queuePoT^2]$ are bounded and equivalent to $\gpdScale / (1-\gpdShape)$ and $2\gpdScale^2 / (1-\gpdShape)(1-2\gpdShape)$, respectively, only if $\gpdShape < 1/2$.
In this regard, 
constraints \eqref{eqn:constraintNW_extremes_bounds} and \eqref{eqn:constraintNW_extremes_bounds_2} for all $\vue\in\vueSet$ can be rewritten as follows:
\begin{gather}
\label{eqn:constraintNW_extremes_bounds_modified}
	\textstyle \lim\limits_{\timeblock \to \infty} \sum_{t=1}^{\timeblock} \big( \queue_{\vue}(t) - \queueTH \big)\indictsimp{\queue_{\vue}(t)} / \sum_{t=1}^{\timeblock} \indictsimp{\queue_{\vue}(t)} \leq \frac{\gpdScale}{1-\gpdShape}, \\
\label{eqn:constraintNW_extremes_bounds_modified_2}
	\textstyle \lim\limits_{\timeblock \to \infty} \sum_{t=1}^{\timeblock} \big( \queue_{\vue}(t) - \queueTH \big)^2 \indictsimp{\queue_{\vue}(t)} / \sum_{t=1}^{\timeblock} \indictsimp{\queue_{\vue}(t)} \leq \frac{2\gpdScale^2}{(1-\gpdShape)(1-2\gpdShape)}.
\end{gather}
Assisted by the RSU, each VUE pair can estimate $\gpdShape$ and $\gpdScale$ locally without sharing its QSI, hence effectively decoupling the constraints \eqref{eqn:constraintNW_extremes_bounds} and \eqref{eqn:constraintNW_extremes_bounds_2}, and imposing them locally as in \eqref{eqn:constraintNW_extremes_bounds_modified} and \eqref{eqn:constraintNW_extremes_bounds_modified_2}, respectively.

\subsection{Lyapunov Optimization for Power Allocation}\label{sec:lyapunov}

By using EVT to model $\queuePoT = \queue-\queueTH (>0)$ and its first two moments, we recast the original problem into an equivalent form:
\begin{subequations}\label{eqn:optimization_NW_modified}
	\begin{eqnarray}
	\underset{[\txpowerVec_{\vue}(t)]_{\forall\vue\in\vueSet}^{\forall t}}{\text{minimize}} && \textstyle \lim_{\timeblock \to \infty}  \frac{1}{\timeblock}\sum_{t=1}^{\timeblock} \left(  \sum_{\vue\in\vueSet} \one\transpose \txpowerVec_{\vue}(t) \right) \\
	\label{eqn:constraint_all}
	\text{subject to} && \eqref{eqn:queue_dynamics}\text{-}\eqref{eqn:queue_reliability}, \eqref{eqn:constraintNW_control_domain}, \eqref{eqn:constraintNW_extremes_bounds_modified}, \eqref{eqn:constraintNW_extremes_bounds_modified_2}.
	\end{eqnarray}
\end{subequations} 

To devise a tractable solution for the modified stochastic optimization problem in \eqref{eqn:optimization_NW_modified}, we resort to Lyapunov optimization \cite{book:neely10}.
To this end, first, we should model the time average constraints as virtual queues.
As such, the reliability constraint in \eqref{eqn:queue_reliability} can be recast as
$\expect[\queue_{\vue}] \leq \outage \queueTH$ for each VUE $\vue\in\vueSet$ using the upper bound condition $\probability( \queue_{\vue} \geq \queueTH) \leq \expect[\queue_{\vue}]  / \queueTH$ based on the Markov's inequality \cite{jnl:Ghosh02}.
Our next goal is to introduce a virtual queue $\vqQ_{\vue}$ for the aforementioned constraint instead of \eqref{eqn:queue_stability} and \eqref{eqn:queue_reliability}.
Now, the time average constraints in \eqref{eqn:constraint_all} for all $\vue\in\vueSet$ are modeled by virtual queues as follows:
\begin{subequations}\label{eqn:virtual_queues_for_constraints}
	\begin{align}
	 \vqQ_{\vue}(t+1) 
	 &= [\vqQ_{\vue}(t) + (\queue_{\vue}(t+1) - \outage \queueTH ) ]^+\\
	\vqEXq_{\vue}(t+1) 
	&= [\vqEXq_{\vue}(t) + 
	\big( \queue_{\vue}(t+1) - 
	\queueTH - \textstyle \frac{\gpdScale}{1-\gpdShape} \big) \indictsimp{\queue_{\vue}(t)} ]^+ \\
	\vqEXqtwo_{\vue}(t+1) 
	&= [\vqEXqtwo_{\vue}(t) +  
	\indictsimp{\queue_{\vue}(t)}\big( \queue_{\vue}(t+1) - \queueTH \big)^2 
	\!\!- \textstyle \frac{2\gpdScale^2 \indictsimp{\queue_{\vue}(t)}}{(1-\gpdShape)(1-2\gpdShape)}   ]^+ 
	\end{align}
\end{subequations}

Let $\queueCombined_{\vue}(t)=[\queue_{\vue}(t),\vqQ_{\vue}(t),\vqEXq_{\vue}(t)),\vqEXqtwo_{\vue}(t))]$ be the combined queue with $\vect{\queueCombined}(t) = [\queueCombined_{\vue}(t)]_{\vue\in\vueSet}$ and its quadratic Lyapunov function $\lyapunov{\vect{\queueCombined}(t)}=\frac{1}{2}\vect{\queueCombined}\transpose(t)\vect{\queueCombined}(t)$.
The one-slot drift of the Lyapunov function is defined as $\lyapunovDrift = \lyapunov{\vect{\queueCombined}(t+1)} - \lyapunov{\vect{\queueCombined}(t)}$.
\begin{proposition}\label{prop:lyapunov_bound_for_drift}
	The upper bound of the Lyapunov drift is given by,
	\begin{multline}\label{eqn:lyapunov_drift}
		\lyapunovDrift 
		\leq 
		\textstyle \sum\limits_{\vue\in\vueSet} \!\Big[	\lyapunovConst 
		+ \big( \arrival_{\vue}(t) - \rate_{\vue}(t) \big) \Big\{ 
		\big( 1 + \vqQ_{\vue}(t) - \outage \queueTH  \big)\queue_{\vue}(t)  \\ 
		- \outage \vqQ_{\vue}(t) 
		+ [  
		2 (\queue_{\vue}(t)-\queueTH) \big( \vqEXqtwo_{\vue}(t) + (\queue_{\vue}(t)-\queueTH)^2 - 
		\thresholdMtwo  \big) \\
		+ \queue_{\vue}(t) 
		+ \vqEXq_{\vue}(t) - \thresholdMone 
		] \indictsimp{\queue_{\vue}(t)} 
		\Big\}
		\Big]
		+ \lyapunovBound,
	\end{multline}
	where $\thresholdMone = \queueTH + \frac{\gpdScale}{1-\gpdShape}$ and $\thresholdMtwo = \frac{2\gpdScale^2}{(1-\gpdShape)(1-2\gpdShape)}$ are the first two moments of the tail distribution of queue lengths.
	A constant bound $\lyapunovBound$ and a set of terms $\{\lyapunovConst\}_{\vue}$ independent from the control variables at time $t$ are given in \eqref{eqn:appndx_lyapunov_bound} and \eqref{eqn:appndx_lyapunov_const}, respectively.
\end{proposition}
\begin{IEEEproof}
	See Appendix \ref{appndx:lyapunov_upperbound}.
\end{IEEEproof}
By controlling the upper bound given in Proposition \ref{prop:lyapunov_bound_for_drift}, the network can ensure the stability of both actual and virtual queues.

The conditional expected Lyapunov drift at time $t$ is defined as $\expect[\lyapunov{\vect{\queueCombined}(t+1)} - \lyapunov{\vect{\queueCombined}(t)} |\vect{\queueCombined}(t) ]$.
We define $\lyapunovTradeoff \geq 0$ as a parameter that controls the tradeoff between the queue length and the accuracy of the optimal solution of \eqref{eqn:optimization_NW_modified}.
We then find the network policies by introducing a penalty term $\lyapunovTradeoff \expect[\sum_{\vue} \one\transpose \txpowerVec_{\vue}|\vect{\queueCombined}(t)]$ to the expected drift and minimizing the upper bound of the drift plus penalty (DPP),
$ \lyapunovTradeoff \expect[\sum_{\vue} \one\transpose \txpowerVec_{\vue}|\vect{\queueCombined}(t)] + \expect[\lyapunovDrift|\vect{\queueCombined}(t)] $.
As a result, our goal will now be to minimize the following upper bound:
\begin{multline}
	\sum_{\vue\in\vueSet} 
	\lyapunovTradeoff \one\transpose \txpowerVec_{\vue}
	+ \big( \arrival_{\vue}(t) - \rate_{\vue}(t) \big) \Big\{ 
	\big( 1 + \vqQ_{\vue}(t) - \outage \queueTH  \big)\queue_{\vue}(t) 
	\\
	+ [  
	\textstyle
	2 (\queue_{\vue}(t)-\queueTH) \big( \vqEXqtwo_{\vue}(t) + (\queue_{\vue}(t)-\queueTH)^2 - 
	\frac{2\gpdScale^2}{(1-\gpdShape)(1-2\gpdShape)}  \big)
	\\
	+ \queue_{\vue}(t)
	+ 
	\vqEXq_{\vue}(t) - \queueTH - \textstyle \frac{\gpdScale}{1-\gpdShape} 
	] \indictsimp{\queue_{\vue}(t)} 
	- \outage \vqQ_{\vue}(t) 
	\Big\},
\end{multline}
at each time $t$.
Assuming that VUEs maintain channel-quality indicators (CQIs), each VUE can estimate the interference $\interference_{\vue}^{\rb}(t) \simeq \interferenceEST_{\vue}^{\rb}(t)$ based on past observations (time averaged interference)~\cite{pap:luoto17}.
Hence, the minimization of the above upper bound can be decoupled among VUEs as follows:
\begin{subequations}\label{eqn:optimization_power_vue}
	\begin{alignat}{2}
	\label{eqn:objective_power_per_vue}
	\underset{\txpowerVec_{\vue}(t)}{\text{minimize}} &  \textstyle \sum\limits_{\rb\in\rbSet_{\zone(t,\vue)}} \!\!\!\! \big[ \lyapunovTradeoff\txpower_{\vue}^{\rb}(t)
	- \si_{\vue}(t) \ln \big( 1 + \cinr_{\vue}^{\rb}(t) \txpower_{\vue}^{\rb}(t) \big) \big] \\
	\label{eqn:constraint_sum_power} \text{subject to} & \quad \textstyle  \sum_{\rb\in\rbSet_{\zone(t,\vue)}} \txpower_{\vue}^{\rb}(t) \leq \txpowerMax, \\
	& \quad \txpower_{\vue}^{\rb}(t) \geq 0 \qquad \forall \rb\in\rbSet_{\zone(t,\vue)},
	\end{alignat}
\end{subequations} 
where 
$\si_{\vue}(t) = \frac{\bandwidth}{\ln 2} \Big\{ 
\big( 1 + \vqQ_{\vue}(t) - \outage \queueTH  \big)\queue_{\vue}(t) 
- \outage \vqQ_{\vue}(t) 
+ [  
\queue_{\vue}(t)
 + 
\vqEXq_{\vue}(t) - \queueTH - \textstyle \frac{\gpdScale}{1-\gpdShape} 
+ 2 (\queue_{\vue}(t)-\queueTH) \big( \vqEXqtwo_{\vue}(t) + (\queue_{\vue}(t)-\queueTH)^2 - 
\frac{2\gpdScale^2}{(1-\gpdShape)(1-2\gpdShape)}  \big)
] \indictsimp{\queue_{\vue}(t)} 
\Big\}$
and 
$\cinr_{\vue}^{\rb}(t) = \frac{ \channel_{\vue\vue}^{\rb}(t)  }{ \interferenceEST_{\vue}^{\rb}(t) + \noise }$.
The optimal solution of the convex optimization problem of \eqref{eqn:optimization_power_vue} is obtained by a \emph{water-filling algorithm} \cite{book:proakis} where  $[\txpower_{\vue}^{\rb}(t)]\optimal = [\frac{\si_{\vue}(t)}{\lyapunovTradeoff + \dualPower\optimal_{\vue}(t)} - \frac{1}{\cinr_{\vue}^{\rb}(t)}]^+$, with $\dualPower_{\vue}(t) \geq 0$ being the Lagrangian dual coefficient corresponding to constraint \eqref{eqn:constraint_sum_power}.
Since the first two moments, $\thresholdMone$ and $\thresholdMtwo$, of the distribution of queue lengths exceeding $\queueTH$ impact the optimal solution $[\txpower_{\vue}^{\rb}(t)]\optimal$, in what follows we propose a mechanism to estimate the GPD parameters accurately.

\section{Learning the Parameters of the Maximum Queue Distribution}\label{sec:federated_learning}

The optimal power allocation problem in \eqref{eqn:optimization_power_vue} relies on the characteristics of the excess queue distribution $\gpd^{\gpdCombined} (\queueMAXvar)$.
Hence, estimating the parameters $\gpdScale$ and $\gpdShape$  with high accuracy using QSI samples gathered over the network is imperative. 
In this regard, modeling the distribution of queue lengths exceeding the threshold requires a central controller (e.g., the RSU) to compute and communicate with all VUEs at each time $t$.

\subsection{Queue Sampling via Block Maxima (BM) }\label{sec:block_maxima}

\begin{figure}[!t]
	\centering
	\includegraphics[width=\myfigfactor\linewidth]{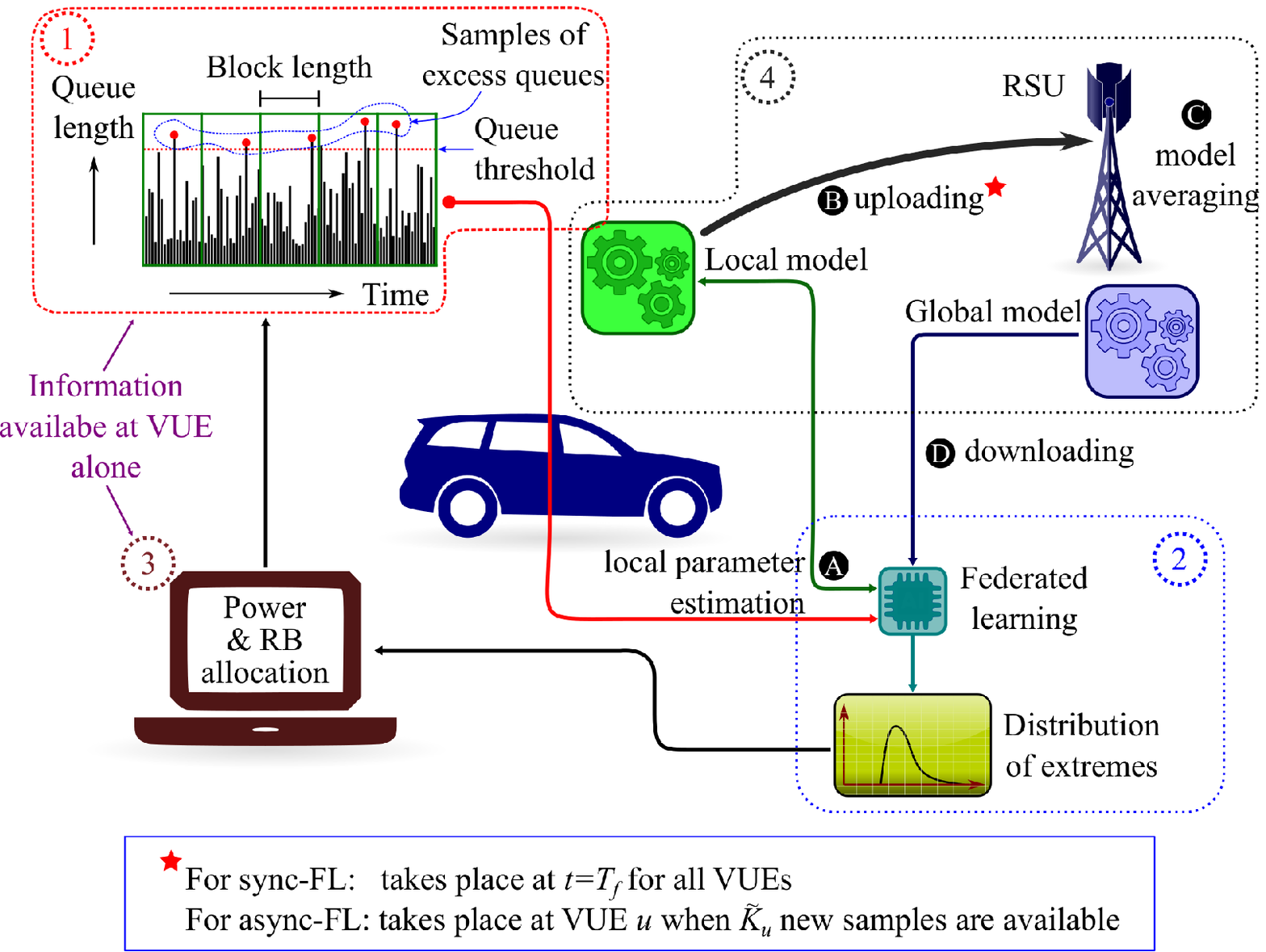}
	\caption{Interrelationships of the processes between VUEs and RSU: 1) excess queue sampling, 2) GPD parameter estimation, 3) transmit power and RB allocation, and 4) local and global models exchange with the RSU.}
	\label{fig:processes_of_vue}
\end{figure}

Let $\blocklength$ be the block length (or time window) during which each VUE draws at most one (the maximum) queue length sample if the queue length exceeds the threshold $\queueTH$.
The size of $\blocklength$ should be sufficiently large to minimize correlation between QSI samples while being sufficiently small to avoid undersampling. 
Henceforth, the assumption of independent queue length samples over VUEs  for EVT-based modeling is satisfied.
We now define $\blockTimeSet_{\block} = \{ (\block-1)\blocklength, (\block-1)\blocklength+1, \dots, \block\blocklength-1 \}$ as the set of time instants during block $\block \in \naturalset$.
Then, the set of queue samples at time $t$ is $\sampleSet_{\vue}(t)= \{\sample_{\vue} = \queue_{\vue}(t_{\block}\optimal) - \queueTH    | \queue_{\vue}(t_{\block}\optimal) > \queueTH , t_{\block}\optimal = \argmax_{\tau\in\blockTimeSet_{\block}} \queue_{\vue}(\tau) , \block \in \{ \seta{ \lfloor t/\blocklength \rfloor } \}  \}$ with a sample size $\BLOCK_{\vue}(t)$.
Note that $0\notin\sampleSet_{\vue}(t)$ for all $\vue\in\vueSet$ and the total number of samples may vary across VUEs since each VUE can independently perform its own QSI sampling process. 
Fig. \ref{fig:processes_of_vue} illustrates each VUE's QSI sampling process.

\subsection{RSU-Centric GPD Parameter Estimation}

As shown in Section \ref{sec:evt}, the distribution of the queue lengths exceeding the threshold is characterized by two parameters $\gpdCombined = [\gpdScale,\gpdShape]$ which need to be accurately estimated.
For this purpose, we use MLE~\cite{book:paul84} whose objective is to find the best set of parameters $\gpdCombined$ that fits the GPD $\gpdML(\cdot)$ to the samples via maximizing the log likelihood function (or minimizing its negative) as follows:
\begin{equation}\label{eqn:MLE_global}
\underset{\gpdCombined\in\gpdCombinedFeasible(\sampleSet)}{\optMinimize} \quad  \loglikelihood(\sampleSet)
= -\frac{1}{\setSize{\sampleSet}}\sum_{\sample\in\sampleSet} \log \gpdML(\sample),
\end{equation} 
where $\gpdCombinedFeasible(\sampleSet) = \{ [\gpdScale,\gpdShape]\in\Re^{2} | \gpdScale > 0, \gpdShape < 1, 1 + \gpdShape \sample /\gpdScale) \geq 0 \text{~for all~} \sample\in\sampleSet \}$ is the feasible set,  and  $\sampleSet = \{\sampleSet_{\vue}\}_{\vue\in\vueSet}$ is the set of network queue length samples.
Here, we omit the time index $t$ for simplicity.
Note that the likelihood function is a smooth function of $\gpdCombined$ and a summation over all the samples in $\sampleSet$, and thus, its gradient over a sample $\sample$ can be derived as follows.

\begin{proposition}\label{prop:derivative_coefficient}
	The derivative coefficient of the negative log-likelihood function of GPD at the queue length sample $\sample$ w.r.t. $\gpdCombined$ is,
	\begin{equation}\label{eqn:gradients}
	\grad{\gpdCombined}{\loglikelihood(\sample)}
	= \begin{bmatrix}
	\daba{\loglikelihood(\sample)}{\gpdScale} \\
	\daba{\loglikelihood(\sample)}{\gpdShape}
	\end{bmatrix}
	= \begin{bmatrix}
	\frac{1}{\gpdScale} \big( \frac{1 + 1/\gpdShape}{ 1 + \gpdShape \sample / \gpdScale } - \frac{1}{\gpdShape} \big)\\
	\frac{(1 + 1/\gpdShape) ( 2 + \gpdShape \sample / \gpdScale )}{ 1 + \gpdShape \sample / \gpdScale } - \frac{ \ln (1 + \gpdShape \sample / \gpdScale) }{\gpdShape^2} 
	\end{bmatrix}.
	\end{equation}
\end{proposition}
\begin{IEEEproof}
	see Appendix \ref{appndx:gevd_gradient}.
\end{IEEEproof}
Using the stochastic variance reduced gradient descent (SVRGD) technique \cite{pap:Johnson13} alongside $\grad{\gpdCombined}{\loglikelihood(\sample)}$ at the RSU, the optimal $\gpdCombined\optimal$ can be derived in an iterative manner (iterating over the sample set) with fast convergence.
For a given predefined step size $\stepsize(>0)$, the evaluation procedure of the GPD parameters using SVRGD over a sample $\sample$ at iteration $\tau$ is defined as follows:
\begin{equation}\label{eqn:svrgd}
\begin{cases}
\vect{y} &= \gpdCombined(\tau) - \stepsize \big[ \grad{\gpdCombined}{\loglikelihoodSP{\gpdCombined(\tau)}}(\sample) - \grad{\gpdCombined}{\loglikelihoodSP{\tilde{\gpdCombined}(\tau)}}(\sample) + \gradient(\tau) \big], \\
\gpdCombined(\tau) &= \argmin_{\gpdCombined\in\gpdCombinedFeasible(\sampleSet)} \| \vect{y} - \gpdCombined  \|, 
\end{cases}
\end{equation}
where $\tilde{\gpdCombined}(\tau) = \frac{1}{\tau-1} \sum_{\tau'=1}^{\tau-1} \gpdCombined(\tau')$ is an average estimate of $\gpdCombined$ over previous iterations and $\gradient(\tau) = \frac{1}{|\sampleSet|} \sum_{\sample\in\sampleSet} \grad{\gpdCombined}{\loglikelihoodSP{\tilde{\gpdCombined}(\tau)}}(\sample)$ is an estimate of the gradient, respectively.
After computing the GPD parameters by iterating over the sample set, RSU shares the optimal GPD parameters with all the VUEs.
This RSU-centric GPD parameter estimation is referred to as ``\CEN{}'', hereinafter, and it is summarized in Algorithm \ref{alg:centralized_estmiation}.

\begin{algorithm}[!t]
	\caption{Centralized GPD parameter estimation in \CEN{}}
	\label{alg:centralized_estmiation}
	\begin{algorithmic}[1]                    
		\STATE \textbf{input:} Gradients $\gradient(0)$, estimations $\gpdCombined(0)$, and step size $\stepsize$ at RSU.
		\FOR{$\federatedTime = 1,2,\ldots$}
		\STATE All VUEs upload their new queue samples $\{\hat{\sampleSet}_{\vue}(\federatedTime)\}_{\vue\in\vueSet}$ to the RSU where $\hat{\sampleSet}_{\vue}(\federatedTime) = \sampleSet_{\vue}(\federatedTime) \setminus \sampleSet_{\vue}(\federatedTime-1)$.
		\STATE \textbf{set:} $\gradient(\federatedTime) = \zero$ and $\gpdCombined(\federatedTime) = \gpdCombined(\federatedTime-1)$ at RSU.
		\STATE Let $\{i^{\block}\}_{\block=1}^{\sum_{\vue}\BLOCK_{\vue}(\federatedTime)}$ be a random permutation of $\{\sampleSet_{\vue}(\federatedTime)\}_{\vue\in\vueSet}$.
		\FOR {$\block = \seta{\sum_{\vue} \BLOCK_{\vue}(\federatedTime)}$}
		\STATE Evaluate $\gradient(\federatedTime)$ and $\gpdCombined(\federatedTime)$ using \eqref{eqn:svrgd} at RSU.
		\ENDFOR
		\STATE Share (download) the GPD parameters $\gpdCombined(\federatedTime)$ with all the VUEs.
		\ENDFOR
	\end{algorithmic}
\end{algorithm}

In \CEN{}, all VUEs in the network need to frequently upload their local queue length samples to the RSU by reusing the RBs available for V2V communication.
This sample uploading over wireless links introduces an additional overhead in which a significant performance degradation can be expected in URLLC.
As the VUE density increases, the sample size grows and so does VUEs' access to the RSU causing congestion resulting in increased network latency.
Henceforth, the need for a distributed learning technique for MLE that does not require sharing all the local samples at VUEs with the RSU or one another is crucial.

\subsection{FL-Based GPD Parameter Estimation}

Towards developing a distributed learning mechanism for GPD parameter estimation, first we rewrite the likelihood function as follows:
\begin{align}\label{eqn:likelihood_local}
	\loglikelihood(\sampleSet)
	= \frac{1}{\setSize{\sampleSet}}\sum_{\sample\in\sampleSet} \log \gpdML(\sample)
	= \sum_{\vue\in\vueSet}\sampleSizeRatio \loglikelihood(\sampleSet_{\vue}),
\end{align}
where $\sampleSizeRatio = \frac{\setSize{\sampleSet_{\vue}}}{\setSize{\sampleSet}} = \frac{\BLOCK_{\vue}}{\sum_{\vue'} \BLOCK_{\vue'}}$.
In \eqref{eqn:likelihood_local}, we express the likelihood function of the network as a weighted sum of likelihood functions per VUE.
Hereinafter, for simplicity, we use $\loglikelihood$ and $\loglikelihood_{\vue}$ instead of $\loglikelihood(\sampleSet)$ and $\loglikelihood(\sampleSet_{\vue})$, respectively.
The idea behind FL is to use $\loglikelihood_{\vue}$ to evaluate $\gpdCombined_{\vue}$ and $\grad{\gpdCombined}{\loglikelihoodSP{\gpdCombined_{\vue}}_{\vue}}$ locally, where $\gpdCombined_{\vue}$ is the local estimate of $\gpdCombined$ at VUE $\vue$, and update the local estimations via sharing the individual learning \emph{models} $( \grad{\gpdCombined}{\loglikelihoodSP{\gpdCombined_{\vue}}_{\vue}}, \gpdCombined_{\vue}, \BLOCK_{\vue}, \sampleMax_{\vue})$ where $\sampleMax_{\vue} = \max \sampleSet_{\vue}$.
Note that sharing $\sampleMax_{\vue}$ over the network through RSU is sufficient to determine the domain $\gpdCombinedFeasible(\sampleSet)$, which is needed for the SVRGD procedure.

To evaluate the gradients and GPD parameters locally, VUE $\vue$ uses the SVRGD with a step size $\stepsize_{\vue} = \stepsize/\BLOCK_{\vue}$ \cite{jnl:jakub16}.
In this case, given the local and global copies of the GPD parameters and gradients at a time $\tau$, $\gpdCombined_{\vue}(\tau)$, $\gpdCombined(\tau)$, $\gradient_{\vue}(\tau)$, and $\gradient(\tau)$, respectively, the local GPD parameters and gradients are updated for each QSI sample $i_{\vue}\in\sampleSet_{\vue}$, process \textbf{A} in Fig. \ref{fig:processes_of_vue}, as follows:
\begin{equation}\label{eqn:local_updates}
\begin{cases}
	\vect{y}_{\vue} = \gpdCombined_{\vue}(\tau) - \stepsize_{\vue} \big[ \grad{\gpdCombined}{\loglikelihoodSP{\gpdCombined_{\vue}(\tau)}_{\vue}}(i_{\vue}) - \grad{\gpdCombined}{\loglikelihoodSP{\gpdCombined(\tau)}_{\vue}}(i_{\vue}) + \gradient(\tau) \big], \\
	\gpdCombined_{\vue}(\tau) = \argmin_{\gpdCombined_{\vue}\in\gpdCombinedFeasible(\sampleSet_{\vue})} \| \vect{y}_{\vue} - \gpdCombined_{\vue}  \|, \\
	\gradient_{\vue}(\tau) = \grad{\gpdCombined}{\loglikelihood_{\vue}}(\tau) + \gradient_{\vue}(\tau).
\end{cases}
\end{equation}
After computing the gradients and GPD parameters locally, each VUE $\vue$ uploads its model at time $\tau$,  $\localmodel_{\tau}$, to the RSU as illustrated in Fig. \ref{fig:processes_of_vue} by process \textbf{B}.

The RSU will then perform model averaging over the network while calculating the global GPD parameters and gradients as per process \textbf{C} in Fig. \ref{fig:processes_of_vue}:
\begin{equation}\label{eqn:global_updates}
\begin{cases}
	\gpdCombined(\tau) = \gpdCombined(\tau-1) + \sum_{\vue} \sampleSizeRatio(\tau-1) \big( \gpdCombined_{\vue}(\tau-1) - \gpdCombined(\tau-1)\big), \\
	\gradient(\tau) = \frac{1}{\sum_{\vue'} \BLOCK_{\vue'}(\tau-1)} \sum_{\vue} \gradient_{\vue}(\tau - 1).
\end{cases}
\end{equation}
Then, the global model $\globalmodel_{\tau}$ is shared with the network (process \textbf{D} in Fig. \ref{fig:processes_of_vue}).

The evaluation and sharing of parameters at the VUEs and the RSU can be done in either a synchronous or asynchronous manner.
In the \emph{synchronous} approach, at the end of a predefined time interval $\federatedTime \gg \blocklength$, all VUEs evaluate their local gradients and simultaneously upload their local models $\localmodel_{\federatedTime}$ to the RSU.
Then the RSU averages out all the local models after which all VUEs download the global model $\globalmodel_{\federatedTime}$.
Here, synchronization may improve the accuracy of the estimation of global gradients.
However, the simultaneous transmissions to the RSU by all VUEs degrades the VUE-RSU data rates and thus, introduces significant delays to ongoing V2V communication.
This synchronous FL approach presented above is dubbed ``\SYNC{}'' hereinafter, and summarized in Algorithm~\ref{alg:federated_mle_sync}.

\begin{algorithm}[!t]
	\caption{MLE for GPD using \SYNC{}}
	\label{alg:federated_mle_sync}
	\begin{algorithmic}[1]                    
		\STATE \textbf{input:} Gradients $\{\gradient_{\vue}(0)\}_{\vue\in\vueSet}$, local estimations $\{\gpdCombined_{\vue}(0)\}_{\vue\in\vueSet}$, and step size $\stepsize$.
		\FOR{$\federatedTime = 1,2,\ldots$}
		\FOR [in parallel] {each VUE $\vue\in\vueSet$} 
		\STATE \textbf{set:} $\gradient_{\vue}(\federatedTime) = \zero$, $\gpdCombined_{\vue}(\federatedTime) = \gpdCombined(\federatedTime)$ and $\stepsize_{\vue} = \stepsize / \BLOCK_{\vue}(\federatedTime)$.
		\STATE Let $\{i^{\block}_{\vue}\}_{\block=1}^{\BLOCK_{\vue}(\federatedTime)}$ be a random permutation of $\sampleSet_{\vue}$.
		\FOR {$\block = \seta{\BLOCK_{\vue}(\federatedTime)}$}
		\STATE Evaluate $\gradient_{\vue}(\federatedTime)$ and $\gpdCombined_{\vue}(\federatedTime)$ using \eqref{eqn:local_updates}.
		\ENDFOR
		\STATE Upload the model $\localmodel_{\federatedTime}$ to RSU.
		\ENDFOR
		\STATE Update $\gpdCombined(\federatedTime)$ and $\gradient(\federatedTime)$ at the RSU using \eqref{eqn:global_updates}.
		\STATE Share (download) the model $\globalmodel_{\federatedTime}$ with all the VUEs.
		\ENDFOR
	\end{algorithmic}
\end{algorithm}

\begin{figure*}[!t]
	\centering
	\includegraphics[width=.7\linewidth]{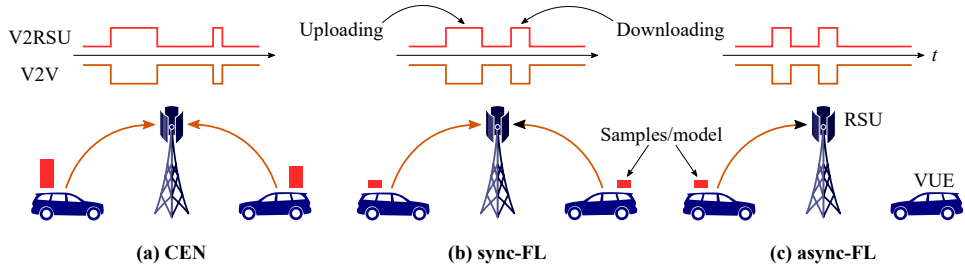}
	\caption{Illustration of QSI/model uploading and downloading for \CEN{}, \SYNC{}, and \ASYNC{}. Switching between VUE-to-RSU (V2RSU) and V2V communications are shown at the top and bottom of the time axis, respectively. }
	\label{fig:cost_communication}
\end{figure*}

In contrast, in the \emph{asynchronous} approach, each VUE must wait until a predefined number $\blockTH$ of new QSI samples are collected.
In essence, at time $\timeFederate_{\vue}$ with $\BLOCK_{\vue}(\timeFederate_{\vue})/\blockTH \in \naturalset$, VUE $\vue$ evaluates and uploads its local model $\localmodel_{\timeFederate_{\vue}}$ to the RSU.
At the RSU, the newly received local model is averaged out with the existing local models of other VUEs, and the updated global model $\globalmodel_{\timeFederate_{\vue}}$ is fed back to VUE $\vue$.
Note that the delay for the upload and download processes will be very small due to the fact that the likelihood of multiple VUEs simultaneously sharing their model is very low.
We designate the asynchronous approach as ``\ASYNC{}'', as seen in Algorithm~\ref{alg:federated_mle_async}.

\begin{algorithm}[!t]
	\caption{MLE for GPD using async-FL}
	\label{alg:federated_mle_async}
	\begin{algorithmic}[1]                    
		\STATE \textbf{input:} Gradients $\{\gradient_{\vue}(0)\}_{\vue\in\vueSet}$, local estimations $\{\gpdCombined_{\vue}(0)\}_{\vue\in\vueSet}$, and step size $\stepsize$.
		\FOR{$t = 1,2,\ldots$}
			\FOR [in parallel] {each VUE $\vue\in\vueSet$} 
				\IF { $\BLOCK_{\vue}(t)/\blockTH \in \naturalset$ }
					\STATE \textbf{set:} $\timeFederate_{\vue} = t$, $\gradient_{\vue}(\timeFederate_{\vue}) = \zero$, $\gpdCombined_{\vue}(\timeFederate_{\vue}) = \gpdCombined(\timeFederate_{\vue})$ and $\stepsize_{\vue} = \stepsize / \BLOCK_{\vue}(\timeFederate_{\vue})$.
					\STATE Let $\{i^{\block}_{\vue}\}_{\block=1}^{\BLOCK_{\vue}(\timeFederate_{\vue})}$ be a random permutation of $\sampleSet_{\vue}$.
					\FOR {$\block = \seta{\BLOCK_{\vue}(\timeFederate_{\vue})}$}
						\STATE Evaluate $\gradient_{\vue}(\timeFederate_{\vue})$ and $\gpdCombined_{\vue}(\timeFederate_{\vue})$ using \eqref{eqn:local_updates}.
					\ENDFOR
					\STATE Upload the model $\localmodel_{\timeFederate_{\vue}}$ to RSU.
				\ENDIF
			\ENDFOR
			\STATE RSU observes the set $\hat{\set{\VUE}}(t)$ of VUEs that upload models at time $t$.
			\IF {$\hat{\set{\VUE}}(t) \neq \emptyset$}
				\STATE Update $\gpdCombined(t)$ and $\gradient(t)$ using \eqref{eqn:global_updates}.
				\STATE Share (download) the model $\globalmodel_{t}$ with all VUEs $\vue \in \hat{\set{\VUE}}(t)$.
			\ENDIF
		\ENDFOR
	\end{algorithmic}
\end{algorithm}

\subsection{Cost of Communication with the RSU}

In all three methods used to estimate the GPD parameters, \CEN{}, \SYNC{}, and \ASYNC{}, the QSI samples or local/global models are exchanged between VUEs and RSU by reusing the RBs available for V2V communication.
This communication between VUEs and the RSU for GPD parameter estimation introduces additional latencies to the ongoing V2V communication.
Such latencies from GPD parameter learning can be seen as additional costs for URLLC applications.
In this regard, modeling the cost of uploading/downloading the learning models or queue samples in terms of an additional delay on V2V communication is illustrated in Fig. \ref{fig:cost_communication} and discussed next.

Let $\sizeGradient$, $\sizeParams$, and $\sizeQueues$ be the sizes (in bits) of gradient values, GPD parameters, and queue samples of any VUE, respectively.
Suppose VUE $\vue$ has $\hat{\BLOCK}_{\vue}$ new samples, and its uplink and downlink rate between RSU are $\rate_{\vue 0}$ and $\rate_{0 \vue}$, respectively.
In the \CEN{} approach, VUE $\vue$ dedicates $\sizeQueues \hat{\BLOCK}_{\vue} / \rate_{\vue 0}$ time to upload all its new queue samples to the RSU while $\sizeParams / \rate_{0 \vue}$ time to download the GPD parameters from the RSU.
Since all VUEs access the RSU simultaneously, the RSU will schedule VUEs over the RBs that are already allocated for their V2V communication links.
As a result, an additional delay of $(\sizeQueues \hat{\BLOCK}_{\vue} / \rate_{\vue 0} + \sizeParams / \rate_{0 \vue})$ is introduced for $\vue$'s V2V communication.

Similar to \CEN{}, in \SYNC{}, the RSU schedules VUEs due to their simultaneous access to the RSU.
However, in \SYNC{}, only the learning models are shared.
Therefore, the corresponding uplink and downlink durations $(\sizeGradient + \sizeParams + \sizeQueues) / \rate_{\vue 0}$ and $(\sizeGradient + \sizeParams + \sizeQueues) / \rate_{0 \vue}$ are introduced as additional delays for VUE $\vue$'s V2V communication.
Similar delays can be observed for \ASYNC{} approach.
However, in \ASYNC{}, VUEs independently access the RSU in which lower interference on VUE-RSU communication links compared to \SYNC{} can be expected.
Therefore, higher rates for $\rate_{\vue 0}$ and $\rate_{0 \vue}$, and lower delays on V2V communication can be expected in \ASYNC{}  compared to the other two methods.

\section{Simulation Results and Analysis}\label{sec:results}

For our simulations, we consider a network based on a 250\,m$\times$250\,m Manhattan mobility model with nine intersections.
In this setting, a road consists of two lanes with 4\,m width in each direction.
We uniformly deploy VUE pairs within each lane with the vRx always following the vTx with a speed of 60\,kmph and a fixed gap of 50\,m.
VUEs share 60 RBs and have a maximum transmit power of $\txpowerMax = 10$\,W.
The RB allocation per \zonedef{} is adopted from \cite{pap:ikram17} and \cite{jnl:liu18}.
The traffic generation at each VUE transmitter follows a Poisson arrival process with a mean of $500\,\text{kbps}$.
For local and global model sharing in FL, the payload sizes of the queue length sample, gradient value, and GPD parameter are assumed to be 8\,bits, 16\,bits, and 16\,bits, respectively.
The rest of the parameter values are presented in Table~\ref{tab:simulation_parameters}.

\begin{table}
	\centering
	\caption{Simulation parameters \cite{pap:ikram16,pap:ikram17,jnl:liu18,pap:mangel11}.}
	\label{tab:simulation_parameters}
	\begin{tabular}{|c c|| c c|| c c|}
		\hline 
		Para. &  Value & Para. &  Value & Para. &  Value \\ 
		\hline \hline
		$\pathlossCoefficient$ & -68.5\,dBm &  $\noiseAlone$ & -174\,dBm/Hz & $\blocklength$ & 10 \\
		$\pathlossCoefficient'$ & -54.5\,dBm & $\stepsize$ & (50,0.005) &$\pathlossExponent$ & 1.61  \\
		$\bandwidth$ & 180\,kHz & $\grad{\gpdCombined}{\loglikelihood_{\vue}}(0)$ & (1,1000) & $\distanceBound $ & 15\,m \\
		$\queueTH$  & 46.29\,kb & $\gpdCombined_{\vue}(0)$ & (1,0) & $\outage$ & 0.001  \\
		\hline 
	\end{tabular} 
\end{table}

\subsection{Centralized vs distributed GPD parameter estimation}

\begin{figure}[!t]
	\centering
	\subfloat[Complementary cumulative distribution functions (CCDF) of queue lengths exceeding $\queueTH$ using \CEN{} and \ASYNC{} methods for different number of VUEs.]{
		\includegraphics[width=\myfigfactorx\linewidth]{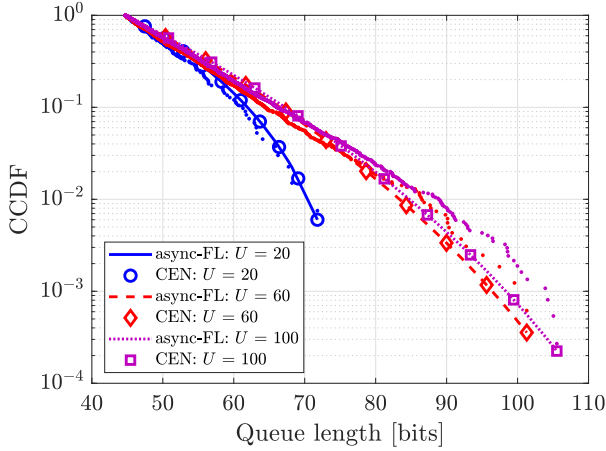}
		\label{fig:test_evt}
	}
\hspace{5pt}
\subfloat[The MLE based cost $\loglikelihood(\sampleSet)$ in \eqref{eqn:MLE_global} as a function of number of iterations used for SVRGD in \CEN{} and \ASYNC{} methods. The selected scenarios are with number of queue length samples about $2\VUE$.]{
	\includegraphics[width=\myfigfactorx\linewidth]{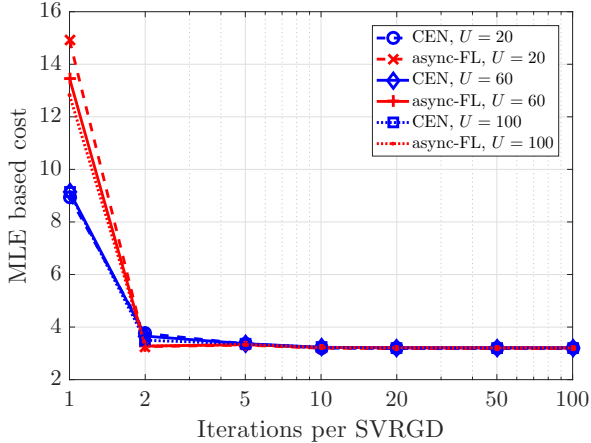}
	\label{fig:mle_cost}
}
\caption{Comparison between \CEN{} and \ASYNC{} in terms of the accuracy of GPD parameter estimation.}	
\label{fig:accuracy}
\end{figure}

Fig. \ref{fig:accuracy} compares the accuracy of GDP parameter estimation using \CEN{} and \ASYNC{}.
In Fig. \ref{fig:test_evt}, the estimated GPDs for \CEN{} and \ASYNC{} approaches for $\VUE=$ 20, 60, and 100 are shown.
Here, the original samples are plotted along the estimated distributions.
From Fig. \ref{fig:test_evt} it can be noted that the estimations of \ASYNC{} are almost equivalent to the \CEN{} estimations.
To evaluate the accuracy of GPD parameter estimation numerically, we use MLE-based cost function in \eqref{eqn:MLE_global} and the corresponding results are illustrated in Fig. \ref{fig:mle_cost}.
Furthermore, in Fig. \ref{fig:mle_cost}, the impact of number of iterations used in SVRGD on the accuracy of GPD parameter estimation in \CEN{} and \ASYNC{} are observed.
Here, the selected scenarios have about $2\VUE$  number of queue length samples exceeding $\queueTH$ in which $\VUE=20$ has the lowest number of samples while $\VUE=100$ has the highest.
When one iteration is used for SVRGD (at the RSU in \CEN{} and at VUEs in \ASYNC{}), Fig. \ref{fig:mle_cost} shows that higher samples yielding lower cost, i.e. better accuracy of GPD parameter estimation.
Therein, the cost of \ASYNC{} is about $66.6\%$, $47.1\%$, and $40.4\%$ higher than of \CEN{} for $\VUE=20$, 40, and 60, respectively.
Increasing the number of iterations used in SVRGD reduces the cost rapidly at first, then the reductions are insignificant.
When two iterations are used for SVRGD, FL yields lower cost and thus, a higher accuracy in parameter estimation compared to \CEN{}.
For larger number of iterations ($>2$) per SVRGD is used, the costs of \ASYNC{} is only about $0.5\%$ higher than the costs of \CEN{} when $\VUE=20$ while for $\VUE=100$, \ASYNC{} yields about $0.5\%$ lower cost compared to \CEN{}.
It highlights that the performance of FL improves over a centralized SVRGD-based estimator with the increasing sample size.

\begin{figure}[!t]
	\centering
	\includegraphics[width=\myfigfactorX\linewidth]{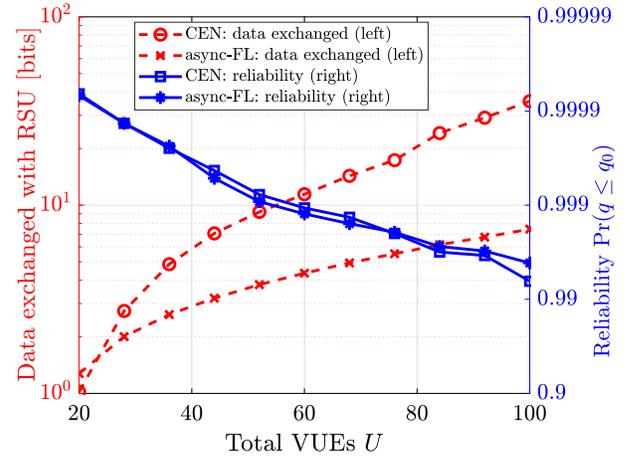}
	\caption{Comparison of the amount of data exchanged between RSU and VUEs (left) and the achieved reliability (right) for different VUE settings with two approaches used to estimate GPD parameters: the proposed FL and the centralized SVRG method.}
	\label{fig:FLvsCEN}
\end{figure}

In Fig. \ref{fig:FLvsCEN}, we compare the amount of data exchange and the achieved reliability in terms of maintaining the queue length below $\queueTH$ for different VUE densities.
As the reliability decreases with increasing the number of VUEs, \ASYNC{} achieves a reliability that is slightly lower to the one resulting from the \CEN{} approach for $\VUE<72$, while outperforms \CEN{} when $\VUE>72$.
Note that the \CEN{} method requires all VUEs to upload all their queue length samples to the RSU and to receive the estimated GPD parameters.
In contrast, in \ASYNC{}, VUEs upload their locally estimated learning model $\localmodel$ and receive the global estimation of the model.
For fewer number of VUEs, $\VUE=20$, the sample size of the network is small, and, thus, \CEN{}  can operate efficiently using very few data samples. 
In contrast, in \ASYNC{}, VUEs must upload and download both parameters and gradients yielding higher data exchange compared to \CEN{}.
However, as the number of VUEs increases (beyond 28), the sample size grows, and thus, \CEN{} incurs higher amount of data exchanged between the RSU and VUEs compared to \ASYNC{}.
The reductions of the exchanged data in \ASYNC{} compared to \CEN{} is about 27\% for $\VUE=28$ and improves up to 79\% when $\VUE=100$.
Finally, Fig. \ref{fig:FLvsCEN} clearly demonstrates that the \ASYNC{} approach is particularly effective for large-scale and dense vehicular networks.

\subsection{Performance Evaluation}

Next, the proposed approaches, \CEN{}, \SYNC{}, and \ASYNC{}, that utilize EVT to characterize the tail distribution of queue lengths are compared with three other baseline models namely: 
\emph{i)} \FP{}:
a V2V network where vTxs use fixed transmit power,
\emph{ii)} \QSO{}:
a V2V network with the objective of power minimization while ensuring only the queue stability \eqref{eqn:queue_dynamics}-\eqref{eqn:queue_stability},
and
\emph{iii)} \QSR{}:
a V2V network that minimizes transmit power while focusing on the probabilistic constraint on average queue length and the queue stability \eqref{eqn:queue_dynamics}-\eqref{eqn:queue_reliability}.

\begin{figure}[!t]
	\centering
	\subfloat[Average transmit power versus number of VUE pairs.]{
		\includegraphics[width=\myfigfactorx\linewidth]{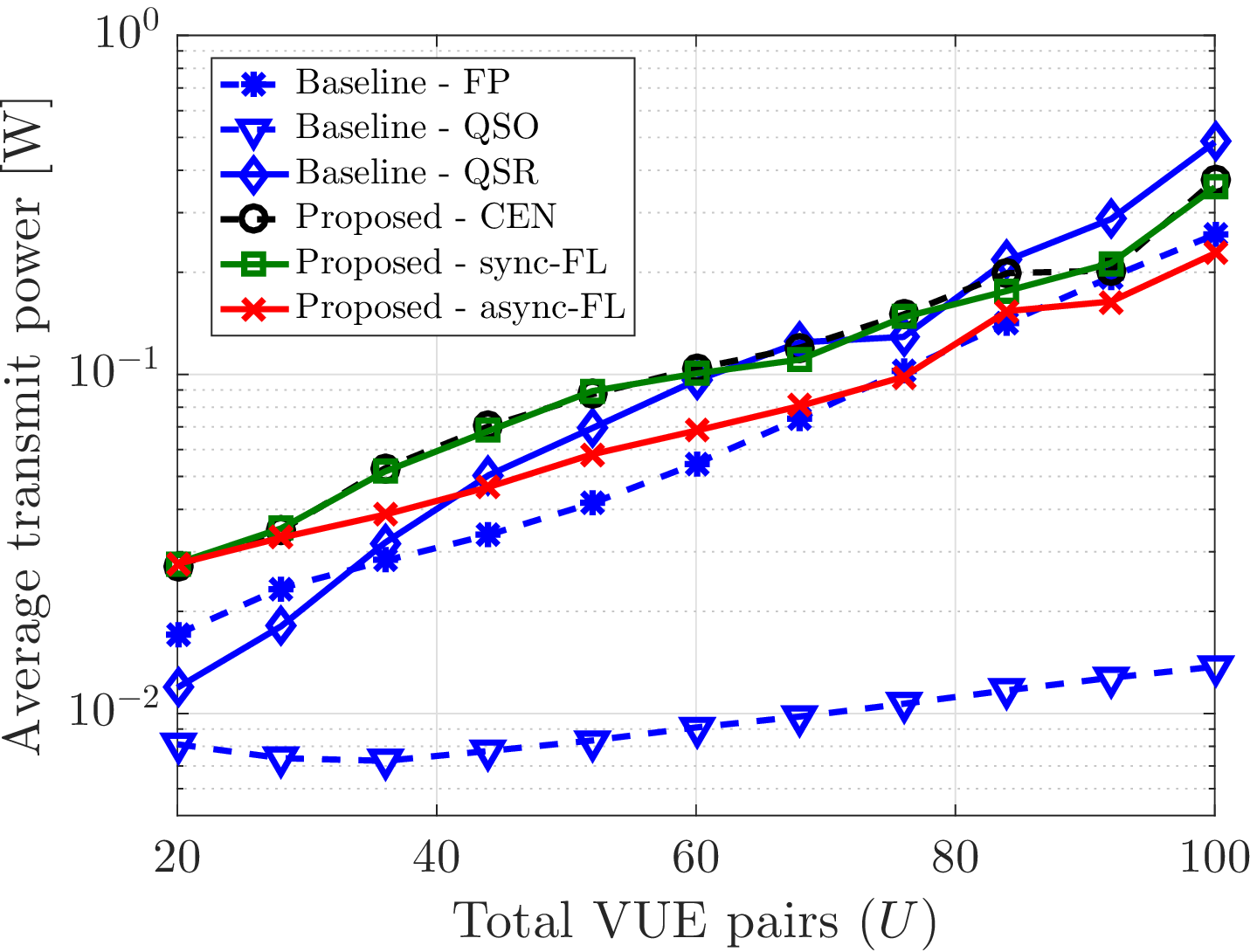}
		\label{fig:power}
	}
\hspace{5pt}
	\subfloat[Average queue length versus number of VUE pairs.]{
		\includegraphics[width=\myfigfactorx\linewidth]{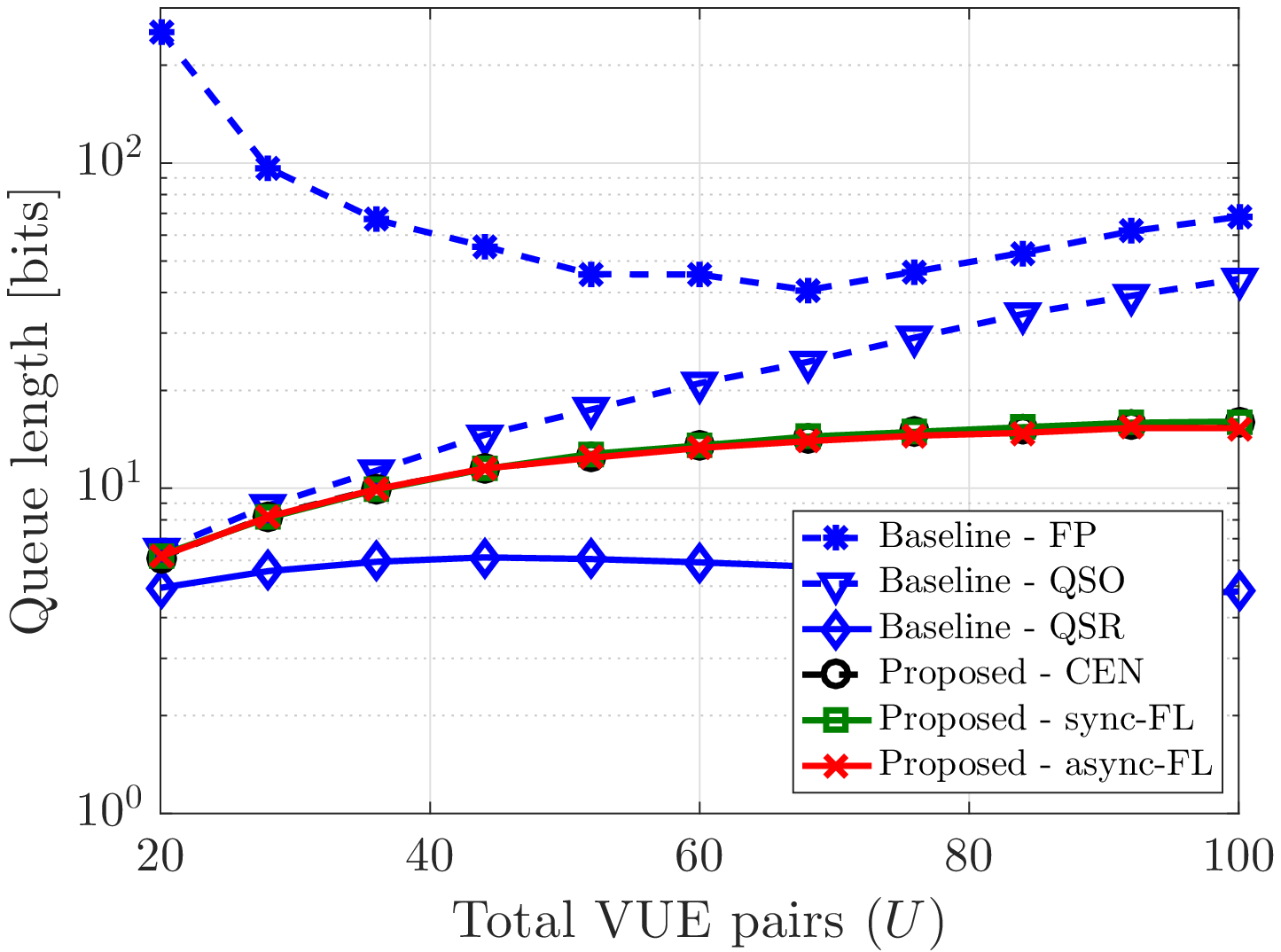}
		\label{fig:queue_avg}
	}
	\caption{Comparison of average transmit power and average queue lengths for different number of VUEs.}
\end{figure}

Fig. \ref{fig:power} compares the average transmit power of all approaches for different VUE densities.
For a fair comparison, the transmit power of vTxs in \FP{} is chosen as the average of transmit powers from all other five methods.
The baseline \QSO{}, which is oblivious to reliability, consumes a minimum transmit power out of all other methods.
\QSR{} baseline takes into account reliability while neglecting VUEs with extreme queue lengths, exhibits lower power consumption compared to all three proposed approaches for total VUEs $\VUE<40$. 
For the cases with $\VUE>40$, \QSR{} consumes higher power compared to \ASYNC{} on average, and beyond $\VUE=80$, it is the most power consuming method.
In \QSR{}, there is no control on the number of VUEs with extreme queue lengths that increases with $\VUE$, and thus, their power consumption degrades the performance of \QSR{}.
Both \CEN{} and \SYNC{} methods exhibit almost equal average power consumption while \ASYNC{} uses less transmission power compared to \CEN{} and \SYNC{}.
The requirement of lower transmit power to upload/download learning models due to asynchronous communication between the RSU and VUEs in \ASYNC{} results the power reductions therein.
The power reductions in \ASYNC{} compared to \CEN{} and \SYNC{} are negligible for $\VUE=20$, improves up to 31.6\% when $\VUE=42$, and remains around 35\% for $\VUE>42$.

The average queue length as a function of total VUE pairs for all baseline and proposed methods are shown in Fig. \ref{fig:queue_avg}.
Here, the average queue length reflects the average queuing latency.
In \FP{}, due to the low and fixed transmit power, the VUE queues grow large even for few VUE pairs.
Since the fixed power is increased with $\VUE$ as shown in Fig. \ref{fig:power}, the average queue length decreases with $\VUE$ first, then rises again due to the increased interference of the network.
Although \QSO{} has the lowest power consumption, it yields higher queuing latency compared to all other methods except \FP{}.
All three proposed techniques exhibit similar queue lengths on average while \QSR{} results in the lowest average queuing latency.
Compared to \QSR{}, all three proposed methods that control VUEs with extreme queue lengths suffer up to three times in average queuing latency when $\VUE=100$.

\begin{figure}[!t]
	\centering
	\subfloat[Maximum queue length corresponding to the worst-case latency.]{
		\includegraphics[width=\myfigfactorx\linewidth]{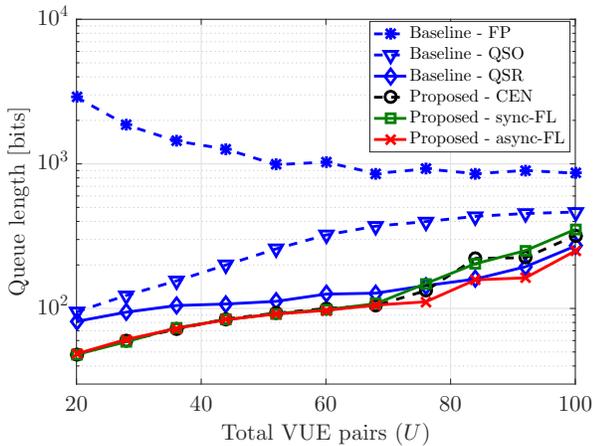}
		\label{fig:queue_max}
	}
	\hspace{5pt}
	\subfloat[Reliability in terms of the probability that the queue lengths are maintained below $\queueTH$.]{
		\includegraphics[width=\myfigfactorx\linewidth]{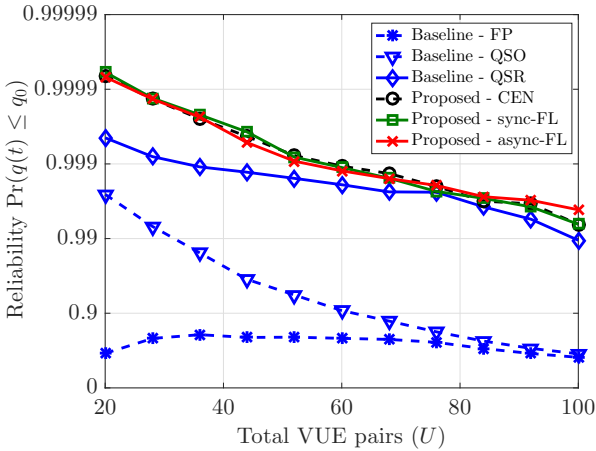}
		\label{fig:queue_reliability}
	}
	\caption{Worst-case latency and reliability for different number of VUEs.}
\end{figure}

Fig. \ref{fig:queue_max} plots the maximum queue length that is proportional to the worst-case latency observed for all methods as a function of the total number of VUE pairs.
Similar to average queue lengths, \FP{} and \QSO{} exhibit the highest worst-case latencies.
\QSR{} which has the lowest average queue lengths displays higher worst-case queue lengths compared to \CEN{} and \SYNC{} for $\VUE\leq 68$ while it fails to outperform \ASYNC{} for all $\VUE \leq 100$.
Although \QSR{} limits the fraction of VUE queue lengths exceeding $\queueTH$ and provides the best average latency, Fig. \ref{fig:queue_max} shows that \QSR{} neglects VUEs with extreme large queue lengths (worst-case VUEs).
All three proposed methods \CEN{}, \SYNC{}, and \ASYNC{} that have control over the tail distribution of the queue lengths yield almost equal worst-case queuing latencies up to $\VUE=68$. 
The reductions in worst-case latencies for all proposed methods are about $40.3\%$ for $\VUE=20$ and about $17.3\%$ for $\VUE=68$ compared to \QSR{}.
Further increasing $\VUE$ increases the number of queue length samples exceeding $\queueTH$ in which frequent communications between VUEs and RSU take place.
As a result, the learning procedure imposes undesirable delays on V2V communication in which high worst-case latencies can be observed in the proposed methods.
However, due to the asynchronous nature of \ASYNC{}, VUEs communicate with the RSU independently in which the delay imposed by model sharing is reduced in \ASYNC{} compared to \CEN{} and \SYNC{}.
Hence, \ASYNC{} yields $7.5\%$, $21.7\%$, and $29\%$ reductions in worst-case latencies compared to \QSR{}, \CEN{}, and \SYNC{}, respectively, for $\VUE=100$.

The reliability in terms of the probability that the queue lengths are maintained below $\queueTH$ for all methods is presented in Fig. \ref{fig:queue_reliability} as a function of the total number of VUE pairs.
It can be noted that \FP{} and \QSO{}, which have no interest in improving V2V communication reliability are the first and second most unreliable methods, respectively.
Since \QSR{} has a reliability constraint, it yields greatly improved reliability over \FP{} and \QSO{}.
\CEN{}, \SYNC{}, and \ASYNC{} control the tail distribution of queue lengths along with the reliability constraint and thus, exhibit further improvements in reliability, i.e. outage reductions, compared to \QSR{}.
Similar to the explanation of the behavior of maximum queue lengths, asynchronous model sharing in \ASYNC{} reduces the delays introduced by the RSU-VUE communications compared to  \CEN{} and \SYNC{} methods.
As a result, for $\VUE \geq 84$, lower queue lengths and thus, reduced outages in \ASYNC{} method over \CEN{} and \SYNC{} can be observed in Fig. \ref{fig:queue_reliability}.
The reductions in outages (or reliability gains) of \ASYNC{} compared to \QSR{} are $84.6\%$ and $18.8\%$ for $\VUE=20$ and 76, respectively.
At $\VUE=100$, \ASYNC{} yields about $60.9\%$, $36\%$, and $35.9\%$ reductions in outages compared to \QSR{}, \CEN{}, and \SYNC{}, respectively.

\begin{figure}[!t]
	\centering
	\includegraphics[width=\myfigfactorX\linewidth]{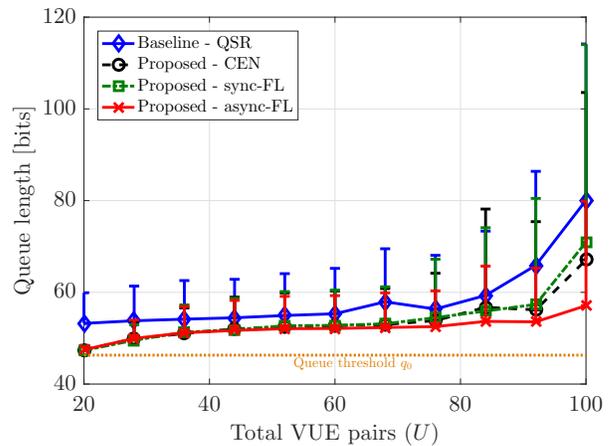}
	\caption{Mean and standard deviation of the tail distribution, i.e., distribution of queue length exceeding $\queueTH$, versus number of VUE pairs.}
	\label{fig:queue_excess}
\end{figure}

\begin{figure}[!t]
	\centering
	\subfloat[CDF of queue lengths comparing \QSR{} and \ASYNC{}.]{
		\includegraphics[width=\myfigfactorx\linewidth]{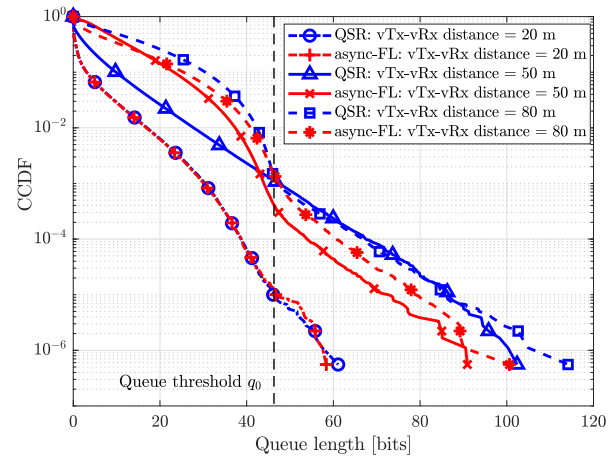}
		\label{fig:dist_q}
	}
	\hspace{5pt}
	\subfloat[CDF of transmit powers comparing \QSR{} and \ASYNC{}.]{
		\includegraphics[width=\myfigfactorx\linewidth]{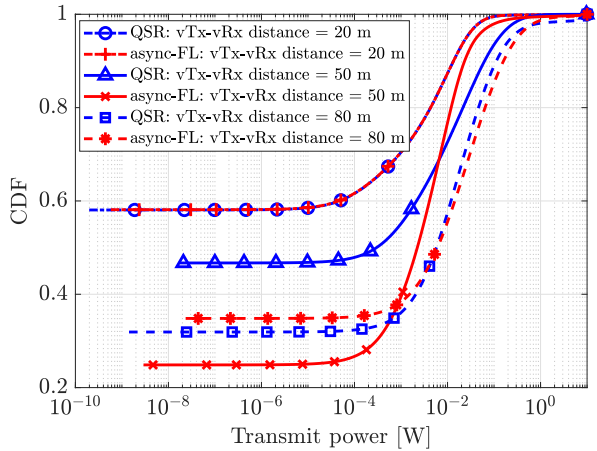}
		\label{fig:dist_power}
	}
	\caption{The impact of vTx-vRx distances on queue lengths and transmit powers.}
	\label{fig:dist}
\end{figure}

\begin{figure*}[!t]
	\centering
	\subfloat[Average queue lengths.]{
		\includegraphics[width=\myfigfactorxx\linewidth]{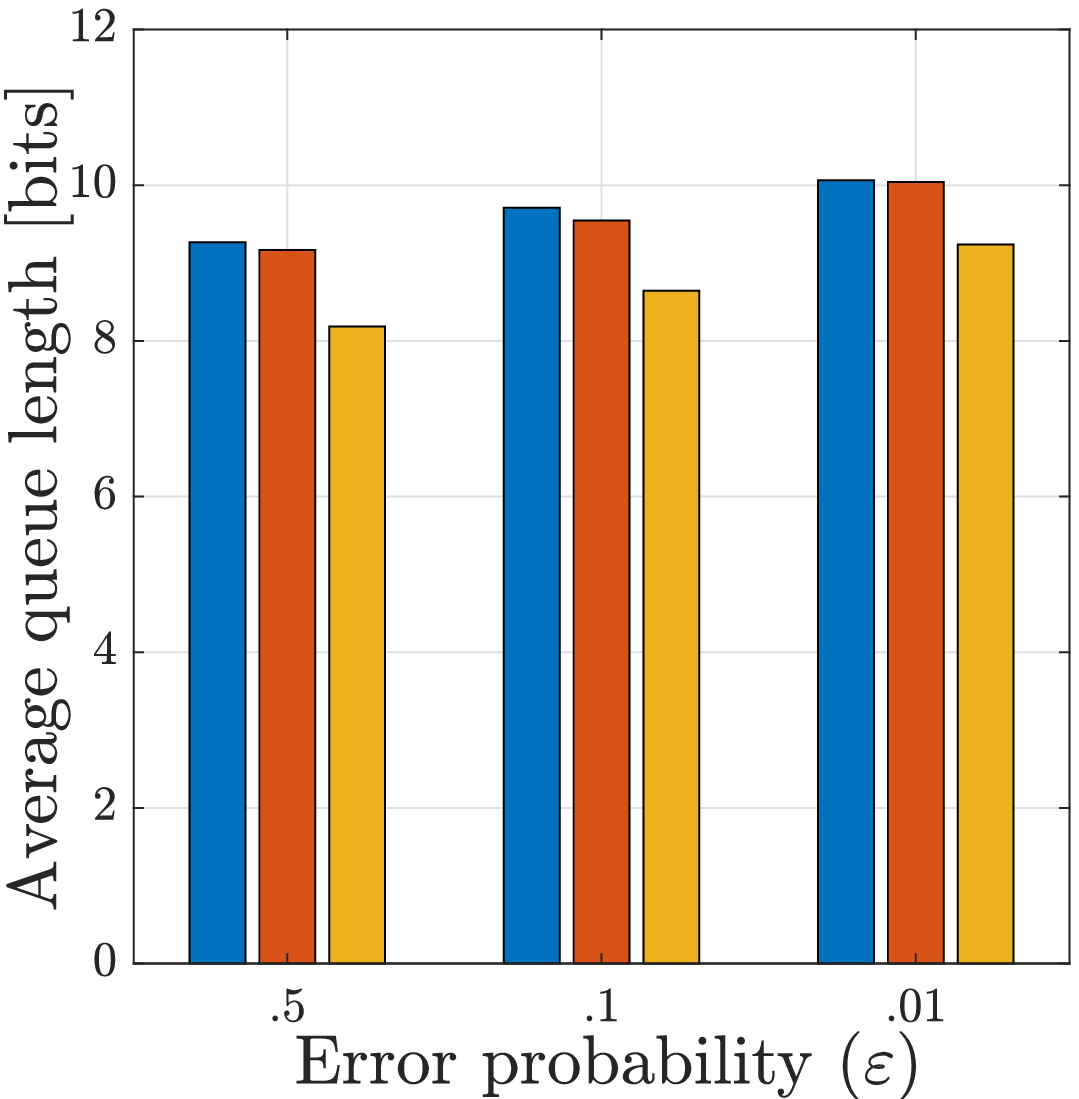}
		\label{fig:fb_Q_avg}
	}
	\subfloat[Maximum queue length.]{
		\includegraphics[width=\myfigfactorxx\linewidth]{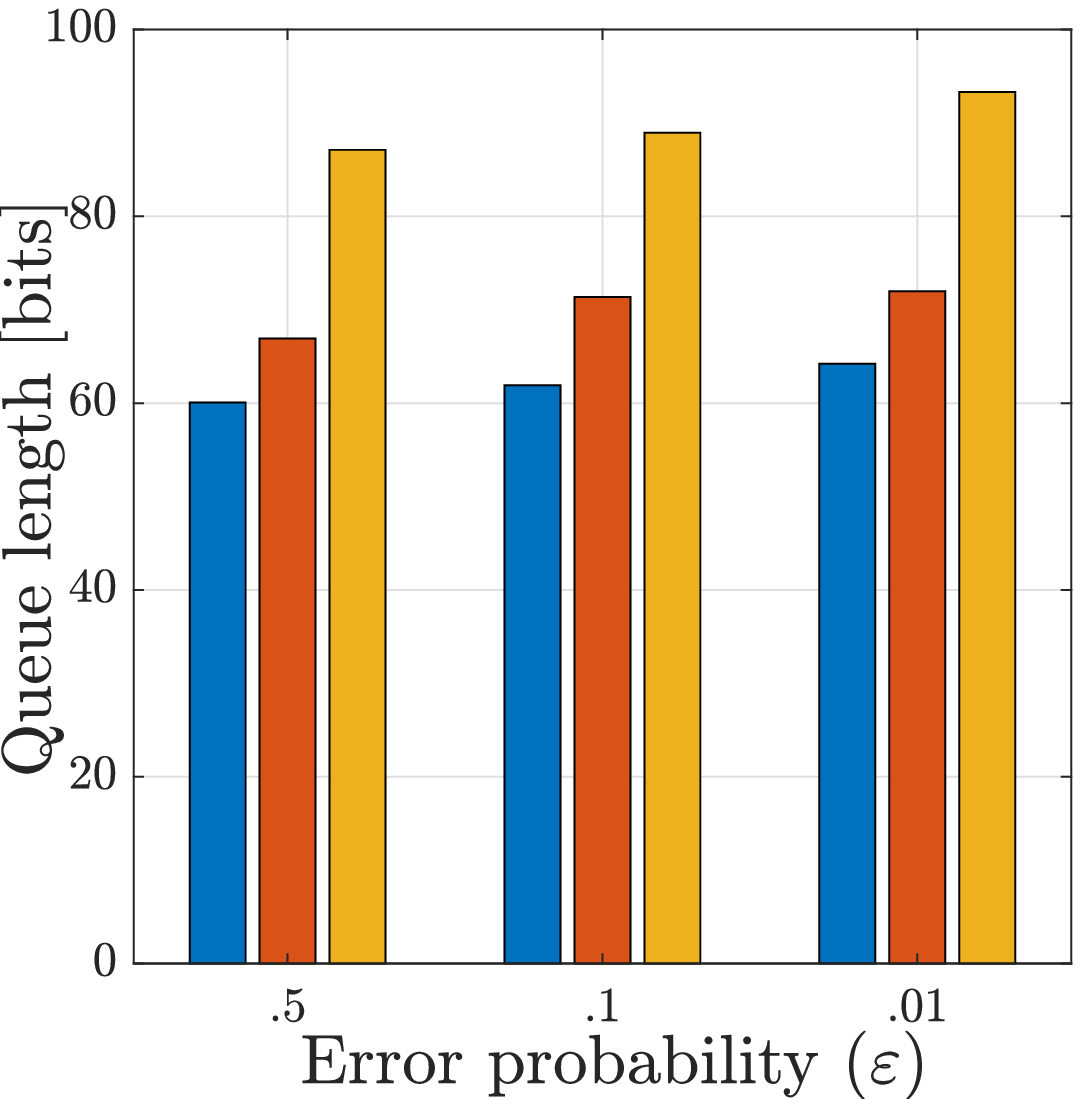}
		\label{fig:fb_Q_max}
	}
	\subfloat[Average transmit power.]{
		\includegraphics[width=\myfigfactorxx\linewidth]{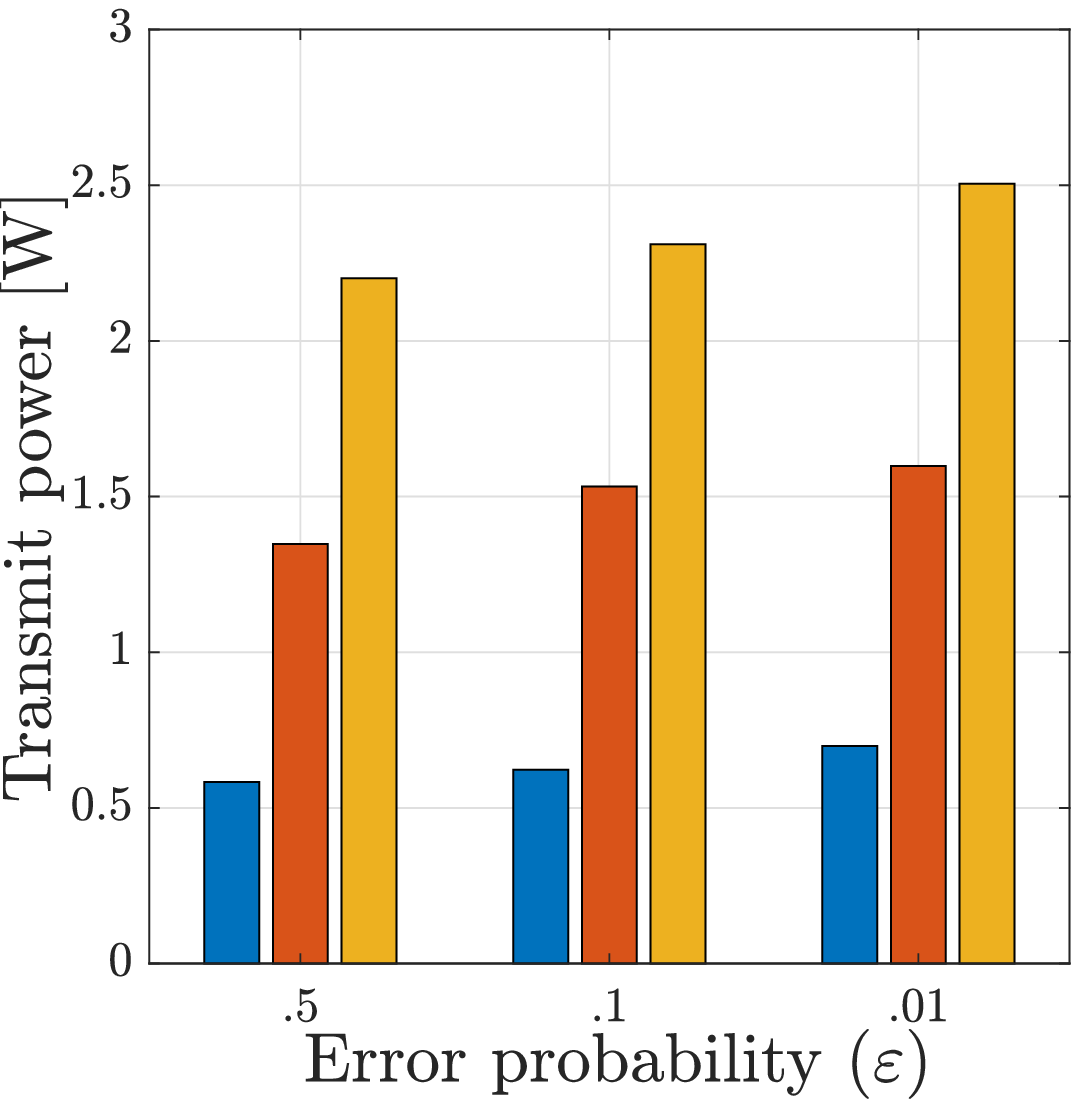}
		\label{fig:fb_P_avg}
	}
	\subfloat[Queue lengths-based outages.]{
		\includegraphics[width=\myfigfactorxx\linewidth]{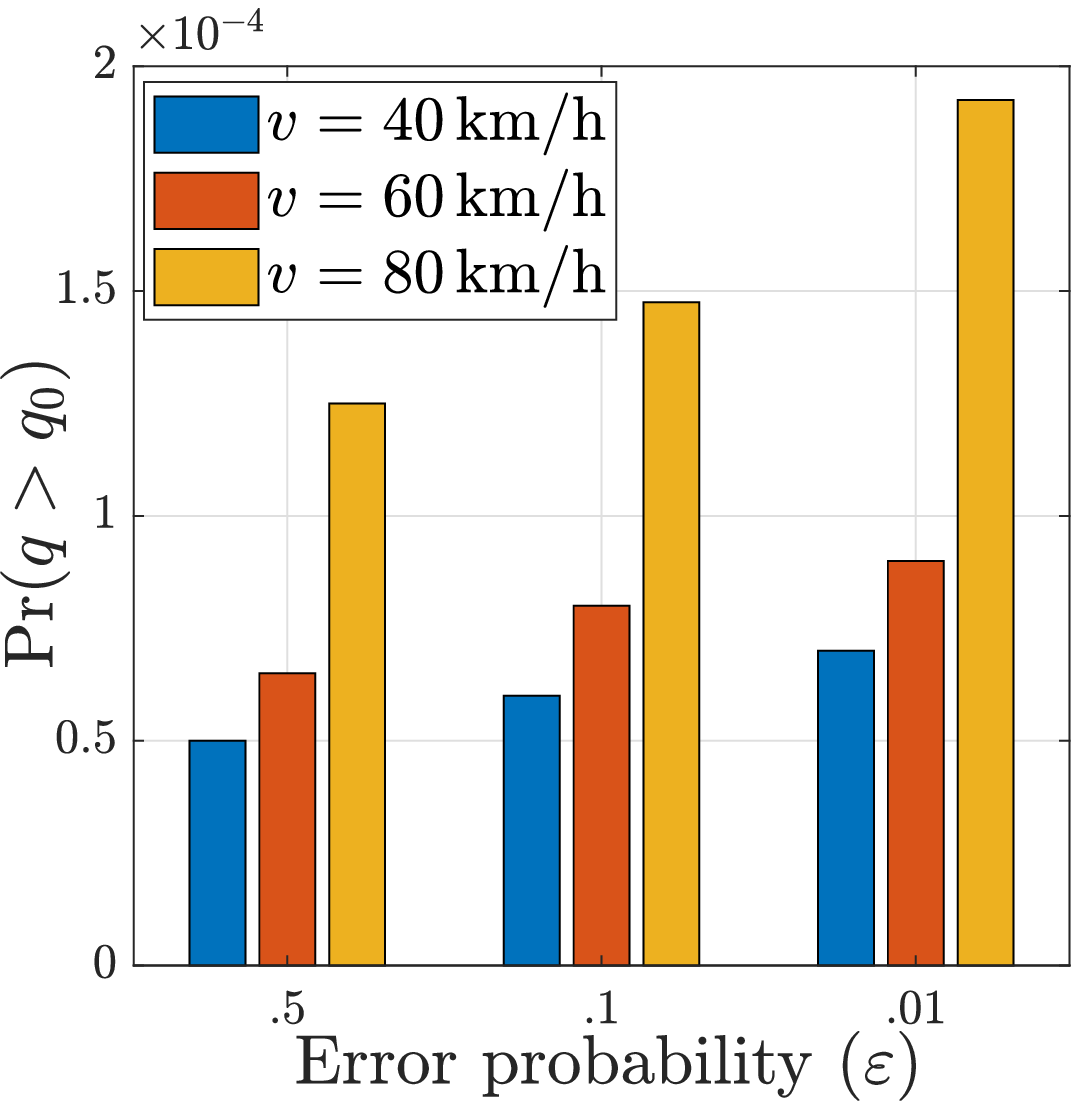}
		\label{fig:fb_REL}
	}
	\caption{Analysis of \ASYNC{} in finite block length regime for different choices of $\fbErr$. Total of 40 VUEs with the speeds of $v=\{40,60,80\}$\,km/h corresponding to the block lengths $\fbLength=\{800,534,400\}$\,bits are considered. }
	\label{fig:fb_results}
\end{figure*}

Fig. \ref{fig:queue_excess} illustrates the mean and standard deviation of the queue length tail distributions of \QSR{}, \CEN{}, \SYNC{}, and \ASYNC{} methods for different number of VUE pairs. 
The standard deviation at a given $\VUE$ is drawn on top of the corresponding mean value to clearly highlight the fluctuations of queue lengths above $\queueTH$.
Note that \FP{} and \QSO{} are neglected since they have large means and standard deviations that do not scale well with the other four methods.
\CEN{} exhibits the lowest means and standard deviations of extreme queue lengths up to $\VUE=44$.
For $\VUE>44$ \ASYNC{} displays the lowest mean and fluctuations of queue lengths exceeding $\queueTH$  proving to be the best candidate for URLLC with large number of VUE pairs.
The reductions in average extreme queue lengths in \ASYNC{} are about $28.6\%$, $41.9\%$, and $19.5\%$ compared to \QSR{}, \CEN{}, and \SYNC{} methods.
From Fig. \ref{fig:queue_excess}, we can see  that \QSR{} has the highest averages of extreme queue lengths compared to \CEN{}, \SYNC{}, and \ASYNC{} methods.
However, the fluctuations of queue lengths above $\queueTH$ are high in \QSR{} only for $\VUE \leq 68$.
Beyond $\VUE=68$, highest fluctuations in extreme queue lengths are seen in both \CEN{} and \SYNC{}.
At $\VUE=100$, the fluctuation reductions in \ASYNC{} are about $33.2\%$, $38\%$, and $47.1\%$ compared to \QSR{}, \CEN{}, and \SYNC{} methods.

The queue length CCDF and transmit power cumulative density function (CDF) of  \QSR{} and \ASYNC{} for different  vTx-vRx distances are shown in Fig. \ref{fig:dist}.
According to Fig. \ref{fig:dist_q}, as vTx-vRx distance increases, the queue lengths increase in both \QSR{} and \ASYNC{} methods due to the reduced over-the-air data rates.
\QSR{} which essentially neglects the queue lengths exceeding $\queueTH$ exhibits longer tails compared to the tail distribution-aware \ASYNC{} method.
In \ASYNC{}, reductions of average queue lengths are $1.2\%$ and $29.4\%$ compared to \QSR{} for vTx-vRx distances of 20\,m and 80\,m, respectively.
For 50\,m, \QSR{} yields  a reduction of $67\%$ in the average queue length over \ASYNC{}.
In terms of reductions of the worst-case queue lengths, VUEs with queue lengths exceeding $\queueTH$, in \ASYNC{} compared to \QSR{} are $1.1\%$, $5.2\%$, and $5.8\%$ for vTx-vRx distances of 20\,m, 50\,m, and 80\,m, respectively.
Fig. \ref{fig:dist_power} shows that both \QSR{} and \ASYNC{} methods consume lower power for the networks with close vTxs and their corresponding vRxs.
For larger vTX-vRx distances, vTxs need higher transmit powers to serve their vRxs, yielding increased transmit powers in both \QSR{} and \ASYNC{} methods.
\ASYNC{} utilizes the characteristics of the queue length tail distribution to reduce the number of VUEs with large queue lengths and thus, minimizes the communications that need high data rates to meet the target reliability.
In contrast, \QSR{} has no control on the queue lengths exceeding $\queueTH$, and thus, requires high data rates to serve VUEs with extreme large queue lengths yielding higher transmit power consumption compared to \ASYNC{}.
Fig. \ref{fig:dist_power} shows that \ASYNC{} method reduces the transmit power consumption on average by $3.2\%$, $18.2\%$, and $43.1\%$ for vTx-vRx distances of 20\,m, 50\,m, and 80\,m, respectively, compared to \QSR{} method.

\subsection{Impact of finite block length}\label{sec:finite_block_length}

	It is worth mentioning that the high mobility in V2V communication networks limits the channel coherence time and thus, the codeword length (or block length) in each transmission.
	The works in \cite{jnl:Xu16,jnl:Hu16,jnl:she17} analyze the URLLC in the finite block length regime and propose resource management solutions therein.
	In this regard, we adopt the approximated achievable rate derived for the finite block length regime from \cite{jnl:she17,jnl:Durisi16} yielding 
	$ \rate_{\vue}
	\approx \bandwidth \Big( \log_2 ( 1 + \sinr_{\vue} ) 
	-  \frac{\sqrt{2\sinr_{\vue}(\sinr_{\vue}+2)} \erfcinv(2\fbErr) }{ \sqrt{\fbLength} (\sinr_{\vue}+1) \ln 2 } \Big)$ where $\sinr_{\vue}$, $\fbLength$, and $\fbErr$ are the signal-to-interference and noise ratio, the block length, and the error probability, respectively.
	Here, the block length $\fbLength$ is proportional to the coherence time, and is inversely proportional to the VUE speed \cite{pap:Pi11}.
	Thus, we consider a total of $\VUE=40$ VUEs with three speeds, \{40, 60, 80\}\,km/h, that correspond to $\fbLength=\{800, 534, 400\}$\,bits.
	The choice of $\fbErr = .5$ represents the original formulation, though the differences in VUE speed are reflected in the channel coherence time, which impacts latency compared with $\fbErr=\{.1, .01\}$.
	Fig. \ref{fig:fb_results} shows the average queue lengths, maximum queue length, average transmit power, and outages in terms of queue lengths exceeding the threshold $\queueTH$ (unreliability) for the \ASYNC{} method.
	For $\fbErr=.5$, which corresponds to the formulation in Section \ref{sec:distributed_solution}, Figs. \ref{fig:fb_Q_max} and \ref{fig:fb_REL} illustrate that both the worst-case queuing latency and the queue outages increase with the VUE speed.
	This is a result of the lower achievable rates due to the smaller coherence times and shorter block lengths at higher speeds.  
	However, higher power consumption yields lower queuing latency on average as exhibited in Fig. \ref{fig:fb_Q_avg}.
	Furthermore, Fig. \ref{fig:fb_results} shows that the performance of the system degrades with decreasing $\fbErr$.

\section{Conclusions}\label{sec:conclusion}
In this paper, we have formulated the problem of joint power control and resource allocation for V2V communication network as a network-wide power minimization problem subject to ultra reliability and low latency constraints.
The constraints in terms of URLLC are characterized using extreme value theory and modeled as the tail distribution of the network-wide queue lengths over a predefined threshold.
Leveraging concepts of federated learning, a distributed learning mechanism is proposed where VUEs estimate the tail distribution locally with the assistance of a RSU.
Here, FL enables VUEs to learn the tail distribution of the network-wide queues locally without sharing the actual queue length samples reducing unnecessary overheads.
Combining both EVT and FL approaches, we have proposed a Lyapunov-based distributed transmit power and resource allocation procedure for VUEs.
Using simulations, we have shown that the proposed method learns the statistics of the network-wide queues with high accuracy.
Furthermore, the proposed method shows considerable gains in reducing extreme events where the queue lengths grow beyond a predefined threshold compared to systems that account for reliability by imposing probabilistic constraints on the average queue lengths.
	It is worth noting that the accuracy of FL-based tail distribution modeling relies on the assumption of IID queue length samples.
	In this regard, a future extension of this work is to study
	FL in the presence of non-IID training data.

\appendices
\section{Proof of Proposition \ref{prop:lyapunov_bound_for_drift}}\label{appndx:lyapunov_upperbound}

First, consider the one-slot drift of the Lyapunov function $\lyapunovDrift 
= \lyapunov{\vect{\queueCombined}(t+1)} - \lyapunov{\vect{\queueCombined}(t)}$.
\begin{multline}
	\lyapunovDrift 
	\label{eqn:appndx_lyapunov_drift}
	= \textstyle \sum\limits_{\vue\in\vueSet} \Big( 
	\frac{ \queue_{\vue}^2(t+1) - \queue_{\vue}^2(t)}{2}
	+ \frac{ \vqQ_{\vue}^2(t+1) - \vqQ_{\vue}^2(t)}{2} \\
	\textstyle + \frac{ \vqEXq_{\vue}^2(t+1) - \vqEXq_{\vue}^2(t)}{2}
	+ \frac{ \vqEXqtwo_{\vue}^2(t+1) - \vqEXqtwo_{\vue}^2(t)}{2}
	\Big).
\end{multline}
Using the relation $([\queue + \arrival - \rate]^+)^2 \leq \queue^2 + (\arrival - \rate)^2 + 2\queue(\arrival - \rate)$ for the $(t+1)$-th terms in \eqref{eqn:appndx_lyapunov_drift}, upper bounds for each of the above terms can be derived as in \eqref{eqn:appndx_all_bounds}.
\begin{figure*}
\begin{subequations}\label{eqn:appndx_all_bounds}
\begin{align}
	\label{eqn:appndx_queue_bound}
	\textstyle
	\frac{ \queue_{\vue}^2(t+1) - \queue_{\vue}^2(t)}{2}
	&\leq \textstyle
	\big[ \frac{( \arrival_{\vue}(t) - \rate_{\vue}(t) )^2}{2} \big]_{\#\text{a}_1} 
	+
	\big[
	{\scriptstyle
		 \queue_{\vue}(t) ( \arrival_{\vue}(t) - \rate_{\vue}(t) )
	}
 	\big], \\
	\nonumber
	\textstyle
	\frac{ \vqQ_{\vue}^2(t+1) - \vqQ_{\vue}^2(t)}{2}
	&\leq \textstyle
	\frac{( \queue_{\vue}(t+1) - \outage \queueTH )^2}{2} 
	+ 
	\big[
	{\scriptstyle
		\vqQ_{\vue}(t) ( \queue_{\vue}(t+1) - \outage \queueTH )
	}\big], \\
	\label{eqn:appndx_vq_reliability}
	&= \textstyle
	\big[ 
	\frac{ \big( \queue_{\vue}^2(t) + ( \arrival_{\vue}(t) - \rate_{\vue}(t) )^2 + \outage^2\queueTH^2 \big)}{2}
	\big]_{\#\text{a}_2}  
	+ \big[ 
	{\scriptstyle
		 \vqQ_{\vue}(t) ( \queue_{\vue}(t) - \outage\queueTH ) - \outage \queueTH \queue_{\vue}(t) 
	}
	\big]_{\#\text{b}_1} 
	+ \big[
	{\scriptstyle
	( \arrival_{\vue}(t) - \rate_{\vue}(t) )
	(\vqQ_{\vue}(t) + \queue_{\vue}(t) - \outage \queueTH)
	}\big], \\
	\nonumber
	\textstyle
	\frac{ \vqEXq_{\vue}^2(t+1) - \vqEXq_{\vue}^2(t)}{2}
	&\leq \textstyle \big\{
	\frac{ \queue_{\vue}^2(t+1) + \thresholdMone^2}{2}
	{\scriptstyle
		- \thresholdMone \vqEXq_{\vue}(t)
	+ ( \vqEXq_{\vue}(t) - \thresholdMone ) \queue_{\vue}(t+1)
	}
	\big\} \indictsimp{\queue_{\vue}(t)}, \\
	\label{eqn:appndx_vq_one}
	&= \textstyle \big\{
	\big[
	{\scriptstyle
		 \frac{ \queue_{\vue}^2(t) + ( \arrival_{\vue}(t) - \rate_{\vue}(t) )^2 + \thresholdMone^2}{2} \big]_{\#\text{a}_3} 
		+ \big[ ( \vqEXq_{\vue}(t) - \thresholdMone ) \queue_{\vue}(t) 
		- \thresholdMone \vqEXq_{\vue}(t)
	} \big]_{\#\text{b}_2} 
	+ 
	\big[
	{\scriptstyle
		( \queue_{\vue}(t) + \vqEXq_{\vue}(t) - \thresholdMone ) ( \arrival_{\vue}(t) - \rate_{\vue}(t) )
	}\big]
	\big\} \indictsimp{\queue_{\vue}(t)}, \\
	\nonumber
	\textstyle
	\frac{ \vqEXqtwo_{\vue}^2(t+1) - \vqEXqtwo_{\vue}^2(t)}{2}
	&\leq \textstyle \big\{
	\frac{ \big( ( \queue_{\vue}(t+1) - \queueTH )^2 - \thresholdMtwo \big)^2}{2}
	{\scriptstyle
		+ \vqEXqtwo_{\vue}(t) \big( ( \queue_{\vue}(t+1) - \queueTH )^2 + \thresholdMtwo \big)
	}
	\big\} \indictsimp{\queue_{\vue}(t)}, \\
	\nonumber
	&= \textstyle \big\{
	\big[ \frac{ \big( ( \queue_{\vue}(t) - \queueTH )^2 + ( \arrival_{\vue}(t) - \rate_{\vue}(t) )^2 - \thresholdMtwo \big)^2}{2}
	{\scriptstyle
	+ 2 \big( ( \queue_{\vue}(t) - \queueTH )^2 + ( \queue_{\vue}(t) - \queueTH ) ( \arrival_{\vue}(t) - \rate_{\vue}(t) ) \big) ( \arrival_{\vue}(t) - \rate_{\vue}(t) )^2
	} \\
	\label{eqn:appndx_vq_two}
	& \qquad 
	{\scriptstyle
		+ \big( ( \queue_{\vue}(t) - \queueTH )^2 + ( \arrival_{\vue}(t) - \rate_{\vue}(t) )^2 \big) \vqEXqtwo_{\vue}(t)
	}
	 \big]_{\#\text{a}_4}  
	+ 
	\big[
	{\scriptstyle 
		2 ( \queue_{\vue}(t) - \queueTH ) ( \arrival_{\vue}(t) - \rate_{\vue}(t) )
		\big( ( \queue_{\vue}(t) - \queueTH )^2 - \thresholdMtwo + \vqEXqtwo_{\vue}(t) \big)
	}\big]
	\big\} \indictsimp{\queue_{\vue}(t)}.
\end{align}
\end{subequations}
\end{figure*}
Here, $\thresholdMone = \queueTH + \gpdScale/(1-\gpdShape)$ and $\thresholdMtwo = 2\gpdScale^2 / (1-\gpdShape)(1-2\gpdShape)$.
Furthermore, along the queue exceeding indicator  $\indictsimp{\queue_{\vue}(t)}$, the definition $\queue_{\vue}(t+1) = \queue_{\vue}(t) + \arrival_{\vue}(t) - \rate_{\vue}(t)$ is used instead of \eqref{eqn:queue_dynamics} due to the fact that $\indictsimp{\queue_{\vue}(t)}=1$ ensures nonempty queues, i.e. $\indictsimp{\queue_{\vue}(t)}=1 \implies \queue_{\vue}(t+1) = [\queue_{\vue}(t) + \arrival_{\vue}(t) - \rate_{\vue}(t)]^+ = \queue_{\vue}(t) + \arrival_{\vue}(t) - \rate_{\vue}(t) > 0$.

Note that the terms $\#\text{a}_1 $-$\#\text{a}_4$ are quadratic in which the assumption of queue stability forces them to be bounded.
Hence sum of $\#\text{a}_1 $-$\#\text{a}_4$ terms are replaced by a bounded value $\lyapunovBound$ given in \eqref{eqn:appndx_lyapunov_bound}.
\begin{figure*}
\begin{multline}\label{eqn:appndx_lyapunov_bound}
	\lyapunovBound \geq 
	\textstyle \sum\limits_{\vue\in\vueSet} \Big[
	\frac{( \arrival_{\vue}(t) - \rate_{\vue}(t) )^2}{2}
	+ 
	\frac{ \big( \queue_{\vue}^2(t) + ( \arrival_{\vue}(t) - \rate_{\vue}(t) )^2 + \outage^2\queueTH^2 \big)}{2}
	+ \big\{
	\frac{ \queue_{\vue}^2(t) + ( \arrival_{\vue}(t) - \rate_{\vue}(t) )^2 + \thresholdMone^2}{2} 
	\textstyle
	+ \frac{ \big( ( \queue_{\vue}(t) - \queueTH )^2 + ( \arrival_{\vue}(t) - \rate_{\vue}(t) )^2 - \thresholdMtwo \big)^2}{2} \\
	+ \big( ( \queue_{\vue}(t) - \queueTH )^2 + ( \arrival_{\vue}(t) - \rate_{\vue}(t) )^2 \big) \vqEXqtwo_{\vue}(t) 
	+ 2 \big( ( \queue_{\vue}(t) - \queueTH )^2 + ( \queue_{\vue}(t) - \queueTH ) ( \arrival_{\vue}(t) - \rate_{\vue}(t) ) \big) ( \arrival_{\vue}(t) - \rate_{\vue}(t) )^2
	\big\} \indictsimp{\queue_{\vue}(t)}
	\Big].
\end{multline}
\end{figure*}
Terms $\#\text{b}_1$ and $\#\text{b}_2$ are independent from the control variables.
Therefore, we denote them by $\lyapunovConst$ for VUE $\vue\in\vueSet$ where
\begin{multline}\label{eqn:appndx_lyapunov_const}
	\lyapunovConst = 
	\vqQ_{\vue}(t) ( \queue_{\vue}(t) - \outage\queueTH ) - \outage \queueTH \queue_{\vue}(t)  \\
	+ ( \vqEXq_{\vue}(t) - \thresholdMone ) \queue_{\vue}(t) 
	- \thresholdMone \vqEXq_{\vue}(t).
\end{multline} 
Combining the results of \eqref{eqn:appndx_all_bounds}-\eqref{eqn:appndx_lyapunov_const} and applying them into \eqref{eqn:appndx_lyapunov_drift} conclude the proof.

\section{Proof of Proposition \ref{prop:derivative_coefficient}}\label{appndx:gevd_gradient}

Let $\gpdExponent{\sample} = (1 + \gpdShape \sample / \gpdScale)^{-1/\gpdShape}$. 
Since $\gpdExponent{\sample} \to e^{- \sample / \gpdScale}$ as $\gpdShape \to 0$, the distribution can be rewritten as $\gpdML(\sample) = \frac{1}{\gpdScale}\gpdExponent{\sample}^{\gpdShape + 1}$.
Using the above notation, it can be noted that, 
\begin{equation}
	\loglikelihood(\sampleSet)
	\textstyle = \sum\limits_{\sample\in\sampleSet} \frac{ \ln \gpdScale - (\gpdShape+1) \ln \gpdExponent{\sample} }{\setSize{\sampleSet}}
	= \frac{1}{\setSize{\sampleSet}}\sum\limits_{\sample\in\sampleSet} \loglikelihood(\sample).
\end{equation}
Hence, $\grad{\gpdCombined}{\loglikelihood(\sampleSet)} = \frac{1}{\setSize{\sampleSet}}\sum_{\sample\in\sampleSet} \grad{\gpdCombined}{\loglikelihood(\sample)}$ is held.

First, the gradient of $\gpdExponent{\sample}$ is found by,
\begin{equation}\label{eqn:gradients_exponent_term}
\grad{\gpdCombined}{\gpdExponent{\sample}}
= \begin{bmatrix}
\frac{\sample}{\gpdScale} \gpdExponent{\sample}^{\gpdShape+1}  \\
\frac{\gpdExponent{\sample} \big( \gpdExponent{\sample}^{\gpdShape} - \ln \gpdExponent{\sample}^{\gpdShape} -1 \big) }{\gpdShape^2}
\end{bmatrix}.
\end{equation}
Thus, the gradient of $\loglikelihood(\sample)$ can be calculated as follows:
\begin{align*}
\grad{\gpdCombined}{\loglikelihood(\sample)}
&= \begin{bmatrix}
\frac{1}{\gpdScale} - \frac{1 + \gpdShape}{\gpdExponent{\sample}} \daba{\gpdExponent{\sample}}{\gpdScale}  \\
- \frac{1 + \gpdShape}{\gpdExponent{\sample}} \daba{\gpdExponent{\sample}}{\gpdShape} - \ln \gpdExponent{\sample}
\end{bmatrix} 
\\
&= \begin{bmatrix}
\frac{1}{\gpdScale} \big( \frac{1 + 1/\gpdShape}{ 1 + \gpdShape \sample / \gpdScale } - \frac{1}{\gpdShape} \big)\\
\frac{(1 + 1/\gpdShape) ( 2 + \gpdShape \sample / \gpdScale )}{ 1 + \gpdShape \sample / \gpdScale } - \frac{ \ln (1 + \gpdShape \sample / \gpdScale) }{\gpdShape^2} 
\end{bmatrix}.
\end{align*}

\bibliographystyle{IEEEtran}
\bibliography{IEEEabrv,mybib_v2x}

\begin{IEEEbiography}[{\includegraphics[width=1in,height=1.25in,clip,keepaspectratio]{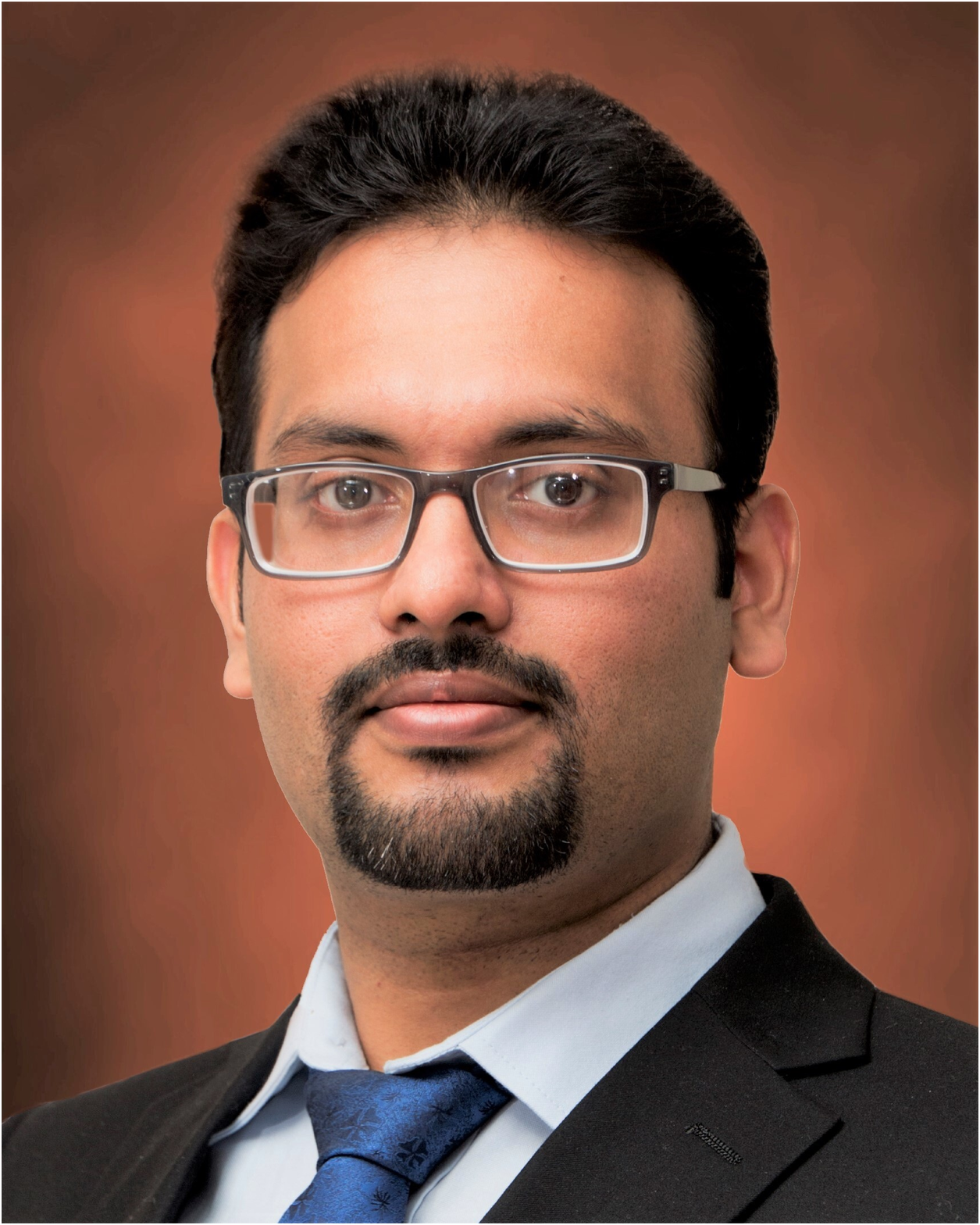}}]{Sumudu Samarakoon}
	(S'08, M'18) received the B.Sc. degree (Hons.) in electronic and telecommunication engineering from the University of Moratuwa, Moratuwa, Sri Lanka, in 2009, the M.Eng. degree from the Asian Institute of Technology, Khlong Luang, Thailand, in 2011, and the Ph.D. degree in communication engineering from the University of Oulu, Oulu, Finland in 2017.
	He is currently with the Intelligent Connectivity and Networks/Systems Group (ICON) in the Centre for Wireless Communications (CWC), University of Oulu, as a Postdoctoral Researcher. 
	His current research interests include heterogeneous networks, small cells, radio resource management, reinforcement learning, and game theory.
	Dr. Samarakoon received the Best Paper Award at the European Wireless Conference and the Excellence Awards for innovators and the outstanding doctoral student at the Radio Technology Unit, CWC, University of Oulu, in 2016. 
\end{IEEEbiography}

\begin{IEEEbiography}[{\includegraphics[width=1in,height=1.25in,clip,keepaspectratio]{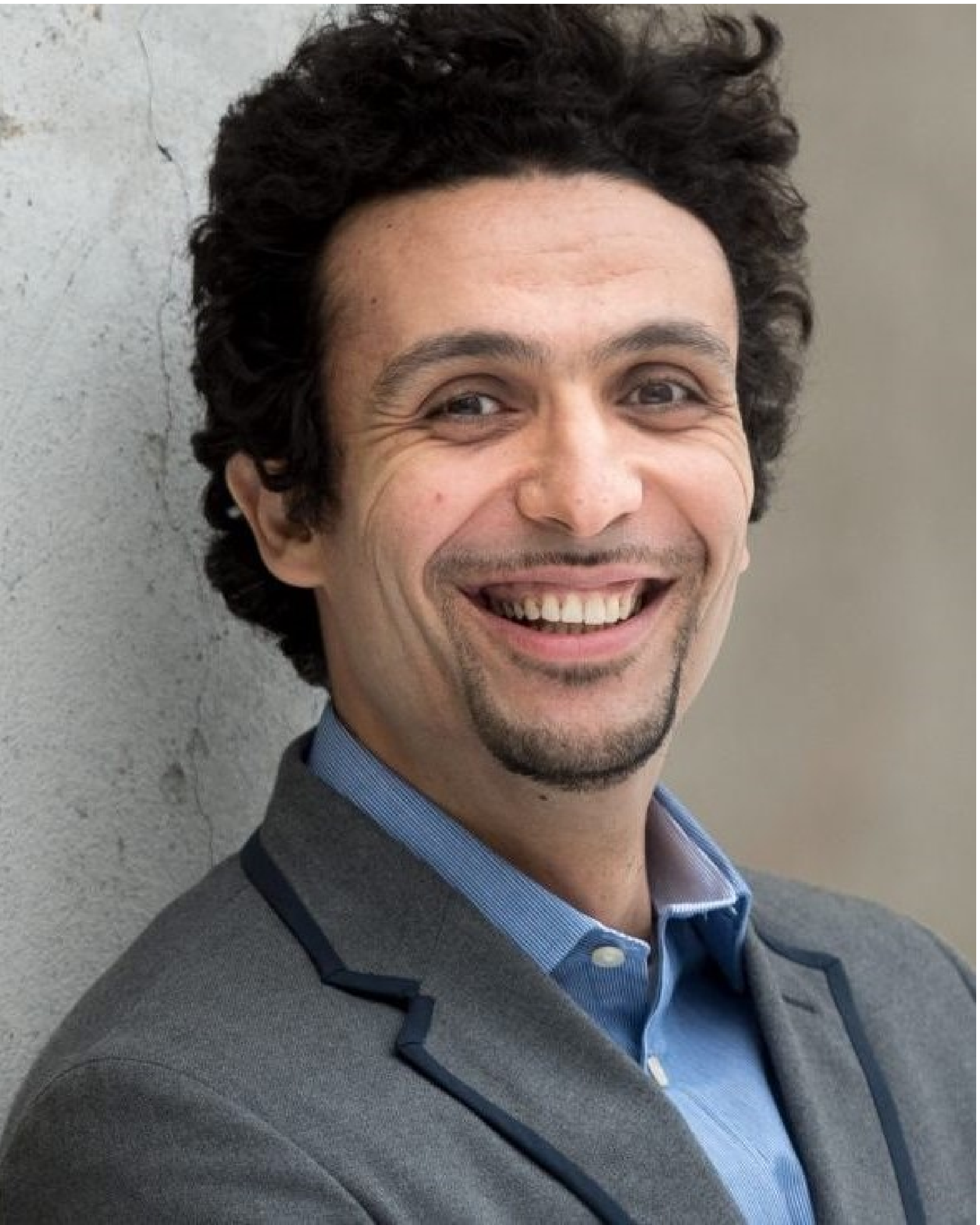}}]{Mehdi Bennis}
	(S'07, AM'08, SM'15)
	is currently an Associate Professor with the Centre for Wireless Communications, University of Oulu, Oulu, Finland, where he is also an Academy of Finland Research Fellow and the Head of the Intelligent Connectivity and Networks/Systems Group (ICON). 
	He has coauthored one book and published more than 200 research articles in international conferences, journals, and book chapters. 
	His current research interests include radio resource management, heterogeneous networks, game theory, and machine learning in 5G networks and beyond.
	Dr. Bennis was a recipient of several prestigious awards, including the 2015 Fred W. Ellersick Prize from the IEEE Communications Society, the 2016 Best Tutorial Prize from the IEEE Communications Society, the 2017 EURASIP Best Paper Award for the Journal on Wireless Communications and Networks, the All-University of Oulu Award for research, and the 2019 IEEE ComSoc Radio Communications Committee Early Achievement Award. 
	He is an Editor of the IEEE Transactions on Communications. 
\end{IEEEbiography}
\vfill

\begin{IEEEbiography}[{\includegraphics[width=1in,height=1.25in,clip,keepaspectratio]{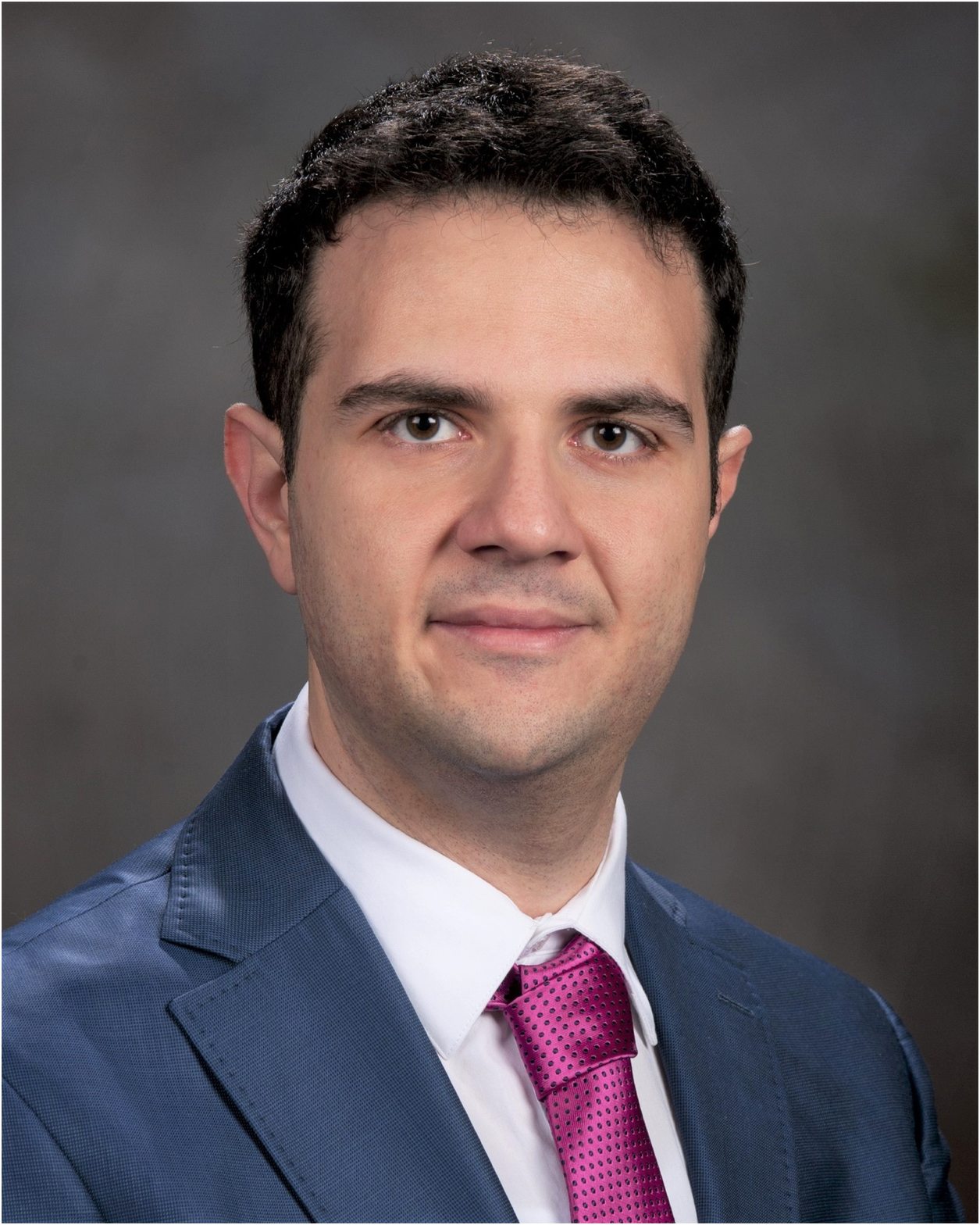}}]{Walid Saad}
	(S'07, M'10, SM'15, F'19) received his Ph.D degree from the University of Oslo in 2010. 
	He is currently a Professor at the Department of Electrical and Computer Engineering at Virginia Tech, where he leads the Network sciEnce, Wireless, and Security (NEWS) laboratory. 
	His research interests include wireless networks, machine learning, game theory, security, unmanned aerial vehicles, cyber-physical systems, and network science.
	Dr. Saad is a Fellow of the IEEE and an IEEE Distinguished Lecturer. He is also the recipient of the NSF CAREER award in 2013, the AFOSR summer faculty fellowship in 2014, and the Young Investigator Award from the Office of Naval Research (ONR) in 2015. 
	He was the author/co-author of eight conference best paper awards at WiOpt in 2009, ICIMP in 2010, IEEE WCNC in 2012, IEEE PIMRC in 2015, IEEE SmartGridComm in 2015, EuCNC in 2017, IEEE GLOBECOM in 2018, and IFIP NTMS in 2019. 
	He is the recipient of the 2015 Fred W. Ellersick Prize from the IEEE Communications Society, of the 2017 IEEE ComSoc Best Young Professional in Academia award, and of the 2018 IEEE ComSoc Radio Communications Committee Early Achievement Award. 
	From 2015-2017, Dr. Saad was named the Stephen O. Lane Junior Faculty Fellow at Virginia Tech and, in 2017, he was named College of Engineering Faculty Fellow. 
	He received the Dean's award for Research Excellence from Virginia Tech in 2019. 
	He currently serves as an editor for the IEEE Transactions on Wireless Communications,  IEEE Transactions on Mobile Computing, IEEE Transactions on Cognitive Communications and Networking, and IEEE Transactions on Information Forensics and Security. 
	He is an Editor-at-Large for the IEEE Transactions on Communications.
\end{IEEEbiography}

\begin{IEEEbiography}[{\includegraphics[width=1in,height=1.25in,clip,keepaspectratio]{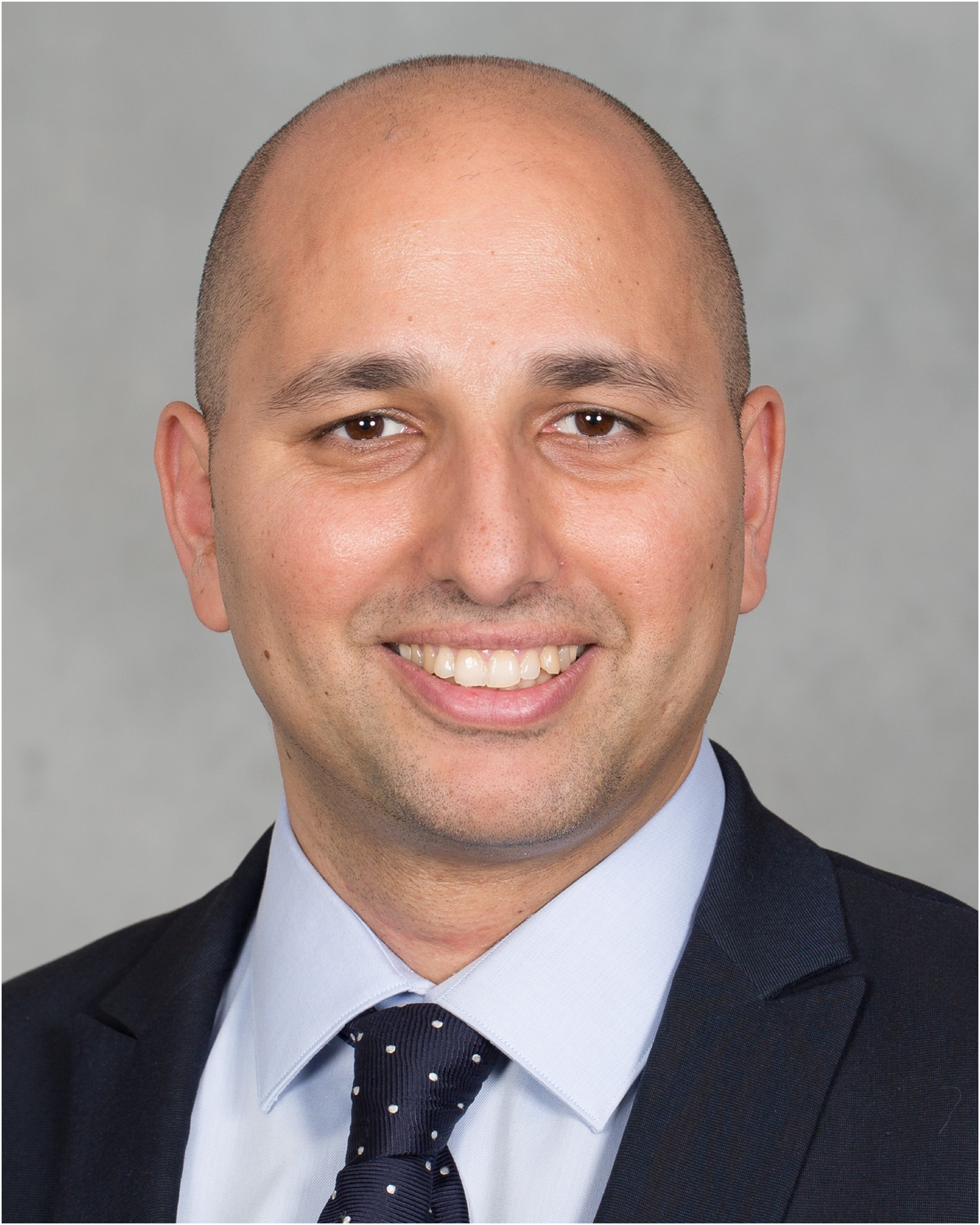}}]{M\'{e}rouane Debbah}
	(S'01, M'04, SM'08, F'15) 
	received the M.Sc. and Ph.D. degrees from the Ecole Normale Sup\'{e}rieure Paris-Saclay, France. 
	He was with Motorola Labs, Saclay, France, from 1999 to 2002, and also with the Vienna Research Center for Telecommunications, Vienna, Austria, until 2003. 
	From 2003 to 2007, he was an Assistant Professor with the Mobile Communications Department, Institut Eurecom, Sophia Antipolis, France. From 2007 to 2014, he was the Director of the Alcatel-Lucent Chair on Flexible Radio. 
	Since 2007, he has been a Full Professor with CentraleSupelec, Gif-sur-Yvette, France.
	Since 2014, he has been a Vice-President of the Huawei France Research Center and the Director of the Mathematical and Algorithmic Sciences Lab. 
	He has managed 8 EU projects and more than 24 national and international projects. 
	His research interests lie in fundamental mathematics, algorithms, statistics, information, and communication sciences research. He is an IEEE Fellow, a WWRF Fellow, and a Membre \'{e}m\'{e}rite SEE. 
	He was a recipient of the ERC Grant MORE (Advanced Mathematical Tools for Complex Network Engineering) from 2012 to 2017. 
	He was a recipient of the Mario Boella Award in 2005, the IEEE Glavieux Prize Award in 2011, and the Qualcomm Innovation Prize Award in 2012. 
	He received 20 best paper awards, among which the 2007 IEEE GLOBECOM Best Paper Award, the Wi-Opt 2009 Best Paper Award, the 2010 Newcom++ Best Paper Award, the WUN CogCom Best Paper 2012 and 2013 Award, the 2014 WCNC Best Paper Award, the 2015 ICC Best Paper Award, the 2015 IEEE Communications Society Leonard G. Abraham Prize, the 2015 IEEE Communications Society Fred W. Ellersick Prize, the 2016 IEEE Communications Society Best Tutorial Paper Award, the 2016 European Wireless Best Paper Award, the 2017 Eurasip Best Paper Award, the 2018 IEEE Marconi Prize Paper Award, the 2019 IEEE Communications Society Young Author Best Paper Award and the Valuetools 2007, Valuetools 2008, CrownCom 2009, Valuetools 2012, SAM 2014, and 2017 IEEE Sweden VT-COM-IT Joint Chapter best student paper awards. 
	He is an Associate Editor-in-Chief of the journal Random Matrix: Theory and Applications. 
	He was an Associate Area Editor and Senior Area Editor of the IEEE TRANSACTIONS ON SIGNAL PROCESSING from 2011 to 2013 and from 2013 to 2014, respectively.
\end{IEEEbiography}
\vfill

\end{document}